\documentclass[preprint]{aastex61}
\usepackage{amsmath}
\usepackage{tabularx}
\usepackage{tabularht}
\usepackage{rotating}
\usepackage[colorinlistoftodos]{todonotes}
\usepackage{hyperref}
\hypersetup{linkcolor=red,citecolor=blue,filecolor=cyan,urlcolor=magenta}

\usepackage[hyphenbreaks]{breakurl}
\usepackage{multirow}
\shorttitle{Comparing LDGRFs and High-Energy SEPs}
\shortauthors{de Nolfo et al.}

\begin{document}

\title{Comparing Long-Duration Gamma-Ray Flares and High-Energy Solar Energetic Particles}


\author{G.~A.~de~Nolfo}\affiliation{Heliophysics Division, NASA Goddard Space Flight Center, Greenbelt, MD, USA.}
\author{A.~Bruno}\affiliation{Heliophysics Division, NASA Goddard Space Flight Center, Greenbelt, MD, USA.}
\author{J.~M.~Ryan}\affiliation{Space Science Center, University of New Hampshire, Durham, NH, USA.}
\author{S.~Dalla}\affiliation{Jeremiah Horrocks Institute, University of Central Lancashire, Preston, PR1 2HE, UK.}
\author{J.~Giacalone}\affiliation{Department of Planetary Sciences, University of Arizona, Tucson, AZ 85721, USA.}
\author{I.~G.~Richardson}\affiliation{Heliophysics Division, NASA Goddard Space Flight Center, Greenbelt, MD, USA.}\affiliation{Department of Astronomy, University of Maryland, College Park, MD, USA.}
\author{E.~R.~Christian}\affiliation{Heliophysics Division, NASA Goddard Space Flight Center, Greenbelt, MD, USA.}
\author{S.~J.~Stochaj}\affiliation{Electrical and Computer Engineering, New Mexico State University, Las Cruces, NM, USA.}
\author{G.~A.~Bazilevskaya}\affiliation{Lebedev Physical Institute, RU-119991 Moscow, Russia.}
\author{M.~Boezio}\affiliation{INFN, Sezione di Trieste, I-34149 Trieste, Italy.}
\author{M.~Martucci}\affiliation{INFN, Sezione di Roma ``Tor Vergata'', I-00133 Rome, Italy.}
\author{V.~V.~Mikhailov}\affiliation{National Research Nuclear University MEPhI, RU-115409 Moscow, Russia.}
\author{R.~Munini}\affiliation{INFN, Sezione di Trieste, I-34149 Trieste, Italy.}


\correspondingauthor{G. A. de Nolfo}
\email{georgia.a.denolfo@nasa.gov}
\correspondingauthor{A. Bruno}
\email{alessandro.bruno-1@nasa.gov}

\date{2019 May 28, accepted for publication in ApJ}

\begin{abstract}
Little is known about the origin of the high-energy and sustained emission from solar Long-Duration Gamma-Ray Flares (LDGRFs), identified with the Compton Gamma Ray Observatory (CGRO), the Solar Maximum Mission (SMM), and now $\it{Fermi}$. Though $\it{Fermi}$/Large Area Space Telescope (LAT) has identified dozens of flares with LDGRF signature, the nature of this phenomenon has been a challenge to explain both due to the extreme energies and long durations. The highest-energy emission has generally been attributed to pion production from the interaction of $\gtrsim$300 MeV protons with the ambient matter. The extended duration suggests that particle acceleration occurs over large volumes extending high in the corona, either from stochastic acceleration within large coronal loops or from back precipitation from coronal mass ejection driven shocks. It is possible to test these models by making direct comparison between the properties of the accelerated ion population producing the $\gamma$-ray emission derived from the $\it{Fermi}$/LAT observations, and the characteristics of solar energetic particles (SEPs) measured by the Payload for Matter-Antimatter Exploration and Light Nuclei Astrophysics (PAMELA) spacecraft in the energy range corresponding to the pion-related emission detected with $\it{Fermi}$. For fourteen of these events we compare the two populations -- SEPs in space and the interacting particles at the Sun -- and discuss the implications in terms of potential sources. Our analysis shows that the two proton numbers are poorly correlated, with their ratio spanning more than five orders of magnitude, suggesting that the back precipitation of shock-acceleration particles is unlikely the source of the LDGRF emission.
\end{abstract}


\section{Introduction}

Among the more unusual solar phenomena are the Long-Duration Gamma-Ray Flares (LDGRFs). The prime feature of these events is delayed and prolonged $\gamma$-ray ($>$100 MeV) emission after the impulsive phase \citep{ref:RYAN2000}. As we discuss below, with the exception of the 2.223 MeV line and other associated emission, there is no clear signal at other wavelengths while this high-energy emission persists. The LDGRF emission is believed to originate from the decay of pions produced by $\gtrsim$300 MeV protons and $\gtrsim$200 MeV $\alpha$ particles. 
Typically, LDGRFs are associated with fast coronal mass ejections (CMEs) and large solar energetic particle (SEP) events, often exceeding the energies of ground level enhancements (GLE, $\gtrsim$500 MeV). However, any direct connection between energetic GLE-type particles observed in space and the accelerated ion population producing the high-energy $\gamma$-ray emission is unclear \citep{ref:CLIVER1989}.
 
During the Solar Maximum Mission (SMM) and the Compton Gamma Ray Observatory (CGRO) missions, only twelve LDGRFs were observed \citep{ref:FORREST1985,ref:RYAN2000,ref:CHUPPRYAN2009}. The LDGRFs of 1991 June observed with CGRO and GAMMA-1 exhibited nuclear emissions for many hours after the impulsive phase \citep{ref:AKIMOV1991,ref:KANBACH1993,ref:RANK2001}. Notably, the 1991 June 11 flare exhibited emission for nearly eleven hours after the impulsive phase with $>$50 MeV emission detected with the Energetic Gamma Ray Experiment Telescope (EGRET) \citep{ref:SCHNEID1994}. \citet{ref:MandzhavidzeRamaty1992} analyzed the EGRET data for the extended phase of the flare. They found a best fit to the emission spectrum with a combination of pion-decay radiation and primary electron bremsstrahlung. However, this was later contraindicated by the Imaging Compton Telescope (COMPTEL) data \citep{ref:RANK2001} showing the 2.223 MeV neutron-capture line synchronized with the $>$100 MeV emission, consistent with a 100$\%$ nuclear origin.
 
More recently, the $\it{Fermi}$/Large Area Telescope (LAT) has observed dozens of LDGRFs \citep{ref:ACKERMANN2012,ref:ACKERMANN2014,ref:ACKERMANN2017,ref:SHARE2018}. These peculiar events share some common characteristics: 1) the extended $\gamma$-ray emission is often delayed by several minutes from the impulsive hard X-ray emission and in some cases lasts for as long as tens of hours; 2) it appears to lack temporal structure on scales much less than the overall decay time, suggesting that the acceleration takes place over large volumes ($\gtrsim10^5$ km), smoothing over the details of the dynamics; and 3) the emission is primarily from high-energy ion interactions, consistent with an origin from $\pi^0$ decay \citep{ref:ACKERMANN2017}. The most intense and longest duration example of an LDGRF is the 2012 March 7 event, for which $>$100 MeV emission was observed for nearly twenty hours \citep{ref:AJELLO2014}. The long duration of the emission cannot be explained by passive trapping without continued production (such as occurs in the Earth's radiation belts), because even the minutest pitch angle scattering would cause particles to escape into the loss cone on much shorter time scales. To produce a 10-hour particle lifetime in a radiation belt-like loop configuration implies scattering mean free paths of tens of AU \citep{ref:RYAN2000} that are not even observed in the most anisotropic GLE events and would require a magnetic fluctuation level $\delta B/B$$\sim$0.05, far quieter than the typical solar wind turbulence. 
 
Several scenarios have been posited to explain the many hours duration of high-energy emission. These include: 
1) particle trapping with and without continuous acceleration within large coronal loops, characterized by delayed onset due to the need for ion energies to exceed the pion production threshold \citep{ref:CHUPPRYAN2009, ref:RYANLEE1991, ref:MandzhavidzeRamaty1992}; 
2) backward precipitation of particles accelerated at a CME-driven shock \citep{ref:CLIVER1993, ref:Kocharov2015}; 
and 3) particle acceleration by large scale electric fields \citep{ref:Litvinenko2006, ref:AKIMOV1991, ref:AKIMOV1996}.
Early models considered the injection of particles into large loops, with a single phase of acceleration followed by precipitation resulting in the observed $\gamma$-ray emission. \citet{ref:MURPHY1987} modeled the emission with two particle populations, while \citet{ref:MandzhavidzeRamaty1992} included the production of $\gamma$-rays from charged pions, the effects of passive particle trapping, and pitch angle scattering into an infinitesimal loss cone. 
Post-flare loops may provide the magnetic structures necessary for continuous acceleration, in which pitch-angle scattering from magnetic turbulence within the loops serves to further accelerate the ion population \citep{ref:RYANLEE1991}. Loops of the necessary length appear during the gradual phase of two-ribbon flares and CME liftoff as field-line reconnection gives rise to hot flare loops that cool within a few hours. The H$\alpha$ loops may be as large as 5$\times$10$^5$ km creating a system of arches that can last for several hours. Such structures may not be visible in soft X rays, however. Recent microwave observations from the Expanded Owens Valley Solar Array (EOVSA) suggest the presence of an extended, static loop, with a circular length of ~1.4 R$_s$ \citep{ref:GARY2018}, for the 2017 September 10 solar flare that produced significant $>$100 MeV $\gamma$-ray emission. 
 
On the other hand, CME-shock-accelerated protons could make their way back from the shock front to the photosphere, radiating $\gamma$-rays, as first proposed by \citet{ref:CLIVER1993}. CME-driven shocks are widely accepted to accelerate the majority of large gradual SEP events \citep{ref:REAMES1999,ref:CLIVER2016} and in so doing transfer typically $\leq$10$\%$ of the CME kinetic energy to energetic particles \citep{ref:MEWALDT2008,ref:ASCHWANDEN2017}. The highest energy events, reaching GeV energies, originate within a few solar radii, where efficient shock acceleration can take place \citep{ref:Zank2000,ref:BerezhkoTaneev2003,ref:LEE2005,ref:Afanasiev2015}. \citet{ref:NGREAMES2008} showed that GeV energies can be reached within $\sim$10 min for a shock speed of 2500 km s$^{-1}$ with typical coronal conditions. Large-scale quiet loops that could efficiently transport particles may connect the shock front (for extended periods) and lead particles back to the photosphere. 
The idea that the same CME-driven shock is responsible for accelerating SEPs and protons producing LDGRFs is supported by recent studies investigating statistical correlations \citep{ref:GOPALSWAMY2018,ref:WINTER2018}.
However, particle diffusion through the turbulent sheath downstream of the shock is expected to be inefficient, and furthermore, the entire 
particle population is being rapidly convected outward. \citet{ref:Kocharov2015} examined the feasibility of back precipitation and could achieve only up to 1$\%$ of GLE particles precipitating back to the Sun. Notably, $\it{Fermi}$/LAT observations revealed three behind-the-limb LDGRFs \citep{ref:ACKERMANN2017}, suggesting a much larger longitudinal extent to the solar source. \citet{ref:PLOTNIKOV2017} investigated the 3-D reconstruction of coronal shocks from these backside events and concluded that LDGRF emission begins when the magnetic connectivity to the shock reaches the solar surface facing the Earth. Indeed, the highest fluxes both in terms of LDGRF emission and in-situ SEPs were observed for the backside event on 2014 September 1, with the fastest moving shock. \citet{ref:JIN2018} simulated this event using a global magnetohydrodynamic (MHD) model and also found that particles could escape downstream of the shock along magnetic field lines that connect to the solar surface facing Earth. 

A CME-driven shock origin is not without challenges. \citet{ref:Hudson2017} pointed out the importance of magnetic mirroring as particles propagate sunwards in preventing back precipitation to the photosphere, consistent with conclusions of \citet{ref:Kocharov2015}. He proposed two alternative scenarios, including the ``Lasso'' scenario, where particles are trapped in a magnetic structure that subsequently retracts back to the solar surface. The second alternative is a coronal ``thick target'' scenario where protons are trapped in a static volume for several hours. Depending on the level of turbulence in this volume, these same particles may be accelerated concommittent with the trapping, extending the potential duration for trapping \citep{ref:RYAN2018a,ref:RYAN2018b}. 
\citet{ref:GRECHNEV2018} found the detectable emission from the backside event of 2014 September 1 to be consistent with flare-accelerated particles trapped in static coronal loops and possibly re-accelerated in these loops by a shock wave excited by the initial eruption. 

High-energy charged particle events detected by in-situ spacecraft at large distances from the Sun and LDGRFs share similar energy ranges for the protons/ions responsible for them. High-energy SEPs and LDGRFs are due to ions rather than electrons, and both are delayed by several minutes from the associated X-ray event \citep{ref:RYAN2000,ref:KAHLER1984}, suggesting a linkage. 
A potential association between LDGRFs and GLE-type particles has been investigated but the results are inconclusive \citep{ref:RAMATYMURPHY1987,ref:RYAN2000,ref:CHUPPRYAN2009,ref:ACKERMANN2017}. 
Absent accompanying signals at other wavelengths, understanding the LDGRF emission is difficult. Only a few options seem to survive scrutiny: 1) LDGRFs contribute directly to SEPs, 2) SEPs produce LDGRFs through back precipitation of particles accelerated at the CME-driven shock, and 3) LDGRFs and high-energy SEP events are correlated but not causally (e.g., the two phenomena may be linked to M and X class flares, but are otherwise separate processes with no exchange of particles). 

One way to constrain the possible scenarios is to compare the number of protons interacting at the Sun above the pion-production threshold ($\sim$300 MeV) 
inferred from the extended $\gamma$-ray emission with the number of SEPs in space above the same energy. This is now possible for the first time with the Payload for Matter-Antimatter Exploration and Light Nuclei Astrophysics (PAMELA) and the accompanying $\gamma$-ray observations with $\it{Fermi}$/LAT. 
In particular, we calculate the total number of $>$500 MeV protons at 1 AU, $N_{SEP}$, taking advantage of the PAMELA and the Solar Terrestrial Relations Observatory (STEREO) data with the aid of transport simulations, and compare it with the number of high-energy protons at the Sun, $N_{LDGRF}$, as deduced from $\it{Fermi}$/LAT data. The ions producing the emission detected with $\it{Fermi}$ are in the same energy range as those observed with PAMELA (see next Section), presenting an opportunity for a proper comparison of the two particle populations, which is key to determining the mechanism responsible for the extended duration of the $\gamma$-ray emission. 

The paper is structured as follows: in Section \ref{PAMELA and Fermi/LAT Observations} we present the SEP observations from PAMELA and LDGRF data from $\it{Fermi}$/LAT for the events used in our analysis; Section \ref{Deriving Total Proton Numbers from SEP Events} describes how we derived information on the spatial extent of SEP events from combined spacecraft data, and how simulations of SEP propagation were used to estimate the number of times particles cross 1 AU; our main results on comparing $N_{LDGRF}$ and $N_{SEP}$ are presented in Section \ref{Results} and discussed in Section \ref{Discussion}; finally, Section \ref{Summary} reports our summary and conclusions.

\section{PAMELA and $\it{Fermi}$/LAT Observations}\label{PAMELA and Fermi/LAT Observations}
PAMELA is a space experiment designed to measure the charged cosmic radiation (protons, electrons, their antiparticles and light nuclei) in the energy range from several tens of MeV up to several hundreds of GeV. The instrument consists of a magnetic spectrometer equipped with a silicon tracking system, a time-of-flight system shielded by an anticoincidence system, an electromagnetic calorimeter and a neutron detector. The Resurs-DK1 satellite, carrying the apparatus, was launched into a semi-polar 70 deg inclination and elliptical (350--610 km) orbit on 2006 June 15. PAMELA provided comprehensive observations of the galactic, solar and magnetospheric radiation in the near-Earth environment \citep{ref:SEP2006,ref:ADRIANI2014,ref:ADRIANI2017,ref:BRUNO_HAWAII,ref:BRUNO_JPCS}. The mission lifetime was extended beyond 2015, in part due to the promise of new PAMELA SEP science, such as that presented in this paper.
On 2016 January 24 the spacecraft lost contact with ground stations. The PAMELA team has recently published the detailed spectra of twenty-six high-energy SEPs between 2006 December and 2014 September \citep{ref:BRUNO2018}. These observations span a broad range in energy from $\sim$80 MeV to 1--2 GeV, encompassing both the low energy measurements of in-situ spacecraft and the ground-based observations of the neutron monitor network. The reported spectra are consistent with diffusive shock acceleration with clear exponential roll-overs attributed to particle escape from within the shock region during acceleration. 
The absence of qualitative differences between the spectra of GLE and non-GLE events suggests that GLEs are not a separate class of SEP events but they rather are the extreme end of a continuous spectral distribution. The PAMELA observations have been used by \citet{ref:BRUNO_GOES} to calibrate the $>$80 MeV proton channels of the Energetic Proton, Electron, and Alpha Detectors (EPEADs) and the High Energy Proton and Alpha Detectors (HEPADs) onboard GOES-13 and -15, bringing the detected spectral intensities in-line with those registered by PAMELA and, thus, enabling a more reliable spectroscopic measurement up to $\sim$1 GeV for SEP events occurring during periods when PAMELA was not acquiring data or after the mission termination \citep{ref:BRUNO2019}.

Fermi/LAT is a pair-conversion telescope with sensitivity to $\gamma$-rays between $\sim$20 MeV and 300 GeV \citep{ref:Atwood2009} and a duty cycle for solar events of only $\sim$15-20\% due to frequent occultation of the Sun by the Earth and passages through South Atlantic Anomaly (SAA). 
However, the $\it{Fermi}$ satellite is able to perform pointed (``target-of-opportunity'') observations increasing the exposure to a particular part of the sky including the Sun.
Dozens of LDGRFs have been reported since the launch of the spacecraft in 2008 June 11 \citep{ref:ACKERMANN2012,ref:ACKERMANN2014,ref:ACKERMANN2017,ref:SHARE2018}.
To derive events from the dataset, a point source is placed at the location of the Sun and a power-law with an exponential cut-off is assumed as the best fit spectral model. All events with photon energies above 100 MeV and directions within 12 deg of the Sun are included in the analysis. 
The $\gamma$-ray background from the Earth's atmosphere is reduced by restricting the allowable events to zenith angles $<$100 deg. 
A solar flux is obtained using a ``maximum likelihood'' 
analysis (\url{https://hesperia.gsfc.nasa.gov/fermi/lat/qlook/max_likelihood/}) that compares the likelihood obtained by fitting the data with the solar source included with the likelihood of the null hypothesis (no solar source).
Details of the analysis of LAT solar flares was published by \citet{ref:ACKERMANN2013}. 
The number of protons inferred from the $\gamma$-ray emission used in this study
rely on the observations of \citet{ref:SHARE2018} based on a ``light bucket''  
approach (\url{https://hesperia.gsfc.nasa.gov/fermi/lat/qlook/light_bucket/}), 
implementing a less accurate but faster algorithm to identify intervals of transient excess high-energy solar emission.
It should be noted that, 
with respect to the ``maximum likelihood'' method, the background is not fitted and the exposure is calculated with an assumed spectral model.

\begin{sidewaystable}[!t]
\center
\footnotesize
\scriptsize
\setlength{\tabcolsep}{8pt}
\renewcommand{\arraystretch}{1.0}
\begin{tabularx}{\linewidth}{c|c|c|c|c|c|c|c|c|c|c}
& \multicolumn{4}{c|}{SEP event} & \multicolumn{3}{c|}{Flare} & \multicolumn{3}{c}{CME}\\
\hline
No & Onset & $>$80 MeV & $>$300 MeV & $>$500 MeV & Onset & Location & Class & 1$^{st}$ app. & Speed & Direction\\
\hline
1 & 2011 03/07, 21:30 & 5.4$\times$10$^{2,\ast}$ & 4.1$\times$10$^{-2,\ast}$ & 3.4$\times$10$^{-4,\ast}$ & 03/07, 19:43 & N30W48 & M3.7 & 03/07, 20:00 & 2223 & N17W50\\
2 & 2011 06/07, 07:00 & (1.5$\pm$0.1)$\times$10$^{5}$ & (3.9$\pm$0.2)$\times$10$^{3}$ & (4.9$\pm$0.8)$\times$10$^{2}$ & 06/07, 06:16 & S21W54 & M2.5 & 06/07, 06:49 & 1321 & S25W52\\
3$\dagger$ & 2011 08/04, 08:05 & (4.9$\pm$1.4)$\times$10$^{4}$ & 2.5$\times$10$^{2,\ast}$ & 1.0$\times$10$^{1,\ast}$ & 08/04, 03:41 & N15W39 & M9.3 & 08/04, 04:12 & 1477 & N14W40\\
4$\dagger$ & 2011 08/09, 08:05 & (2.8$\pm$0.5)$\times$10$^{4}$ & (5.1$\pm$2.0)$\times$10$^{2}$ & (4.5$\pm$4.3)$\times$10$^{1}$ & 08/09, 07:48 & N17W69 & X6.9 & 08/09, 08:12 & 1640 & S12W62\\
5 & 2011 09/06, 23:35 & (1.9$\pm$0.1)$\times$10$^{4}$ & (7.0$\pm$1.0)$\times$10$^{2}$ & (1.0$\pm$0.3)$\times$10$^{2}$ & 09/06, 22:12 & N14W18 & X2.1 & 09/06, 23:05 & 830 & N20W20\\
6 & 2012 01/23, 04:20 & (1.5$\pm$0.2)$\times$10$^{5}$ & (1.3$\pm$0.2)$\times$10$^{2}$ & (6.1$\pm$3.1)$\times$10$^{0}$ & 01/23, 03:38 & N28W21 & M8.8 & 01/23, 04:00 & 2511 & N41W26\\
7 & 2012 01/27, 18:40 & (5.6$\pm$0.4)$\times$10$^{5}$ & (1.1$\pm$0.1)$\times$10$^{4}$ & (2.0$\pm$0.2)$\times$10$^{3}$ & 01/27, 17:37 & N27W71 & X1.8 & 01/27, 18:27 & 2541 & N40W75\\
\multirow{4}{*}{8} & \multirow{4}{*}{2012 03/07, 01:40} & \multirow{4}{*}{(5.2$\pm$1.1)$\times$10$^{6}$} & \multirow{4}{*}{(8.7$\pm$3.0)$\times$10$^{4}$} & \multirow{4}{*}{(1.1$\pm$0.4)$\times$10$^{4}$} & 03/07, 00:02 & N17E27 & X5.4 & 03/07, 00:24 & 3146 & N30E60\\
& & & & & 03/07, 01:05 & N22E12 & X1.3 & 03/07, 01:30 & 2160 & N04E23\\
& & & & & 03/09, 03:22 & N16W02 & M6.3 & 03/09, 04:26 & 1229 & N08E25\\
& & & & & 03/10, 17:15 & N18W26 & M8.4 & 03/10, 18:00 & 1638 & N22E05\\
9 & 2012 05/17, 01:50 & (2.9$\pm$0.1)$\times$10$^{5}$ & (2.0$\pm$0.1)$\times$10$^{4}$ & (5.3$\pm$0.4)$\times$10$^{3}$ & 05/17, 01:25 & N11W76 & M5.1 & 05/17, 01:48 & 1596 & S10W75\\
10 & 2012 07/07, 00:05 & (1.5$\pm$0.1)$\times$10$^{4}$ & (2.0$\pm$0.8)$\times$10$^{2}$ & (1.6$\pm$1.4)$\times$10$^{1}$ & 07/06, 23:01 & S13W59 & X1.1 & 07/06, 23:24 & 1907 & S35W65\\
11 & 2013 04/11, 08:00 & (1.0$\pm$0.1)$\times$10$^{5}$ & (1.4$\pm$0.2)$\times$10$^{3}$ & (8.9$\pm$3.1)$\times$10$^{1}$ & 04/11, 06:55 & N09E12 & M6.5 & 04/11, 07:24 & 1369 & S07E25\\
12 & 2013 05/13, 16:30? & 7.3$\times$10$^{1,\ast}$ & 7.3$\times$10$^{-6,\ast}$ & \nodata & 05/13, 15:48 & N10E89 & X2.8 & 05/13, 16:08 & 1852 & N10E70\\
13 & 2013 05/15, 02:00? & 2.0$\times$10$^{2,\ast}$ & 2.0$\times$10$^{-2,\ast}$ & 5.5$\times$10$^{-4,\ast}$ & 05/15, 01:25 & N11E63 & X1.2 & 05/15, 01:48 & 1408 & N15E70\\
14 & 2013 10/28, 17:55 & (1.7$\pm$0.1)$\times$10$^{4}$ & (5.4$\pm$0.9)$\times$10$^{2}$ & (7.4$\pm$2.5)$\times$10$^{1}$ & 10/28, 15:07 & S06E28 & M4.4 & 10/28, 15:36 & 1098 & N10E20\\
15 & 2014 01/06, 08:05 & (1.0$\pm$0.0)$\times$10$^{5}$ & (5.0$\pm$0.2)$\times$10$^{3}$ & (9.4$\pm$0.6)$\times$10$^{2}$ & 01/06, 07:30\tablenotemark{a} & S15W112\tablenotemark{b} & $\sim$X3.5\tablenotemark{b} & 01/06, 08:00 & 1431 & S03W102\\
16 & 2014 01/07, 19:20 & 2.2$\times$10$^{5,\ast}$ & 1.7$\times$10$^{3,\ast}$ & 1.9$\times$10$^{2,\ast}$ & 01/07, 18:04 & S15W11 & X1.2 & 01/07, 18:24 & 2246 & S24W30\\
17 & 2014 02/25, 03:00 & (1.3$\pm$0.1)$\times$10$^{5}$ & (4.2$\pm$0.2)$\times$10$^{3}$ & (8.1$\pm$1.0)$\times$10$^{2}$ & 02/25, 00:39 & S12E82 & X5.0 & 02/25, 01:25 & 2153 & S11E78\\
18 & 2014 09/01, 17:00 & (2.1$\pm$0.1)$\times$10$^{5}$ & (9.7$\pm$0.7)$\times$10$^{3}$ & (1.5$\pm$0.6)$\times$10$^{3}$ & 09/01, 10:54\tablenotemark{c} & N14E127\tablenotemark{b} & $\sim$X2.4\tablenotemark{b} & 09/01, 11:12 & 2017 & N01E155\\
19 & 2014 09/10, 19:45 & (6.0$\pm$0.4)$\times$10$^{4}$ & (1.0$\pm$0.1)$\times$10$^{3}$ & 1.0$\times$10$^{2,\ast}$ & 09/10, 17:21 & N14E02 & X1.7 & 09/10, 18:00 & 1652 & N15W10\\
20$\dagger$ & 2015 06/21, 02:30? & (7.7$\pm$4.7)$\times$10$^{2}$ & 5.9$\times$10$^{-1,\ast}$ & 3.4$\times$10$^{-2,\ast}$ & 06/21, 02:06 & N13E12 & M2.6 & 06/21, 02:36 & 1740 & N07E08\\
21$\dagger$ & 2015 06/25, 09:30 & 4.4$\times$10$^{1,\ast}$ & 1.6$\times$10$^{-3,\ast}$ & 2.9$\times$10$^{-5,\ast}$ & 06/25, 08:02 & N09W42 & M7.9 & 06/25, 08:36 & 1805 & N23W42\\
22$\dagger$ & 2017 09/06, 12:20 & (1.8$\pm$0.4)$\times$10$^{4}$ & (1.3$\pm$0.6)$\times$10$^{2}$ & (1.9$\pm$1.0)$\times$10$^{1}$ & 09/06, 11:53 & S08W33 & X9.3 & 09/06, 12:24 & 1571\tablenotemark{$\natural$} & S15W23\\
23$\dagger$ & 2017 09/10, 16:05 & (4.5$\pm$1.3)$\times$10$^{6}$ & (1.1$\pm$0.3)$\times$10$^{5}$ & (2.3$\pm$0.6)$\times$10$^{4}$ & 09/10, 15:35 & S08W88 & X8.2 & 09/10, 16:00 & 3163\tablenotemark{$\natural$} & S12W85 \\
\hline
\end{tabularx}
\footnotesize{
(\textit{$\dagger$}) PAMELA data not available. (\textit{$\ast$}) Upper limit. (\textit{$\natural$}) Projected speed (space speed not available).
}
\tablerefs{\footnotesize{
(\textit{a}) \citet{ref:THAKUR2014},
(\textit{b}) \citet{ref:ACKERMANN2017},
(\textit{c}) \citet{ref:PLOTNIKOV2017}.
} 
}
\caption{List of SEP events with an associated LDGRF detected by $\it{Fermi}$/LAT above 100 MeV between 2008 August and 2017 September, based on \citep{ref:SHARE2018,ref:WINTER2018}.
The first five columns report the SEP event number, onset time (UT) and the event-integrated intensities (sr$^{-1}$cm$^{-2}$) above 80, 300 and 500 MeV. 
The question marks indicate uncertain onset times.
For the four eruptions occurring on 2012 March (\#8) a single SEP intensity value is provided. 
Columns 6--9 indicate the parent flare onset time (UT), location (deg) and class, based on GOES soft X-ray data.
Subsequent three columns report the associated CME first-appearance time (UT) and space speed (km s$^{-1}$) according to the CDAW catalog, and direction (deg) from the DONKI database. The dots (...) indicate no data available. See the text for details.}
\label{tab:sep-events}
\end{sidewaystable}

Table \ref{tab:sep-events} reports the SEP events 
with an associated LDGRF detected by $\it{Fermi}$/LAT above 100 MeV between 2008 August and 2019 April. The LDGRF list  
is based on \citet{ref:SHARE2018} except for events $\#$15, $\#$16, $\#$19 and $\#$20, identified with the maximum likelihood analysis 
by \citet{ref:WINTER2018},
and the two events occurring in 2017 ($\#$22 and $\#$23), derived from the light-bucket list of LAT observations (\url{https://hesperia.gsfc.nasa.gov/fermi/lat/qlook/lat_events.txt}).
The first five columns provide the event number, the SEP event onset times (UT) and the event-integrated intensities of protons above 80, 300 and 500 MeV, based on the extrapolation of fits of the PAMELA spectra \citep{ref:BRUNO2018}; data from GOES-13/15 are used below 80 MeV to constrain the fits at low energies, based on the mean energies provided by \citet{ref:SANDBERG2014}. For events not registered by PAMELA (the two events on 2011 August, the 2012 March 7 event, and those occurring after 2014), the fits are based on the EPEAD and HEPAD spectral points as described in \citet{ref:BRUNO2019}, using the mean energies derived by \citet{ref:BRUNO_GOES}. The uncertainties on the event-integrated intensities are computed from the covariance matrix of the fits. 
For the four eruptions occurring on 2012 March (\#8) a single SEP intensity value is provided. Columns 6--9 list the parent flare onset time (UT), location (deg) and soft X-ray class from the GOES X-ray archive (\url{ftp://ftp.ngdc.noaa.gov/STP/space-weather/solar-data/solar-features/solar-flares/x-rays/goes/}). 
For the 2014 January 6 and September 1 events, originating on the far-side of the Sun, the flare size was estimated by \citet{ref:ACKERMANN2017} from observations made by the Extreme Ultraviolet Imager (EUVI) onboard the STEREO spacecraft 
Columns 9--11 provide the associated CME first appearance time (UT) and 
space (3-D) velocity (km s$^{-1}$) according to the Coordinated
Data Analysis Workshops (CDAW; \url{https://cdaw.gsfc.nasa.gov/CME_list/halo/}) catalog of the Large Angle and Spectrometric Coronagraph (LASCO) onboard the Solar and Heliospheric Observatory (SOHO), and the CME direction from the Database Of Notifications, Knowledge, Information (DONKI; \url{https://kauai.ccmc.gsfc.nasa.gov/DONKI/}); for the two 2017 September CMEs the space speeds are not available, so the corresponding projected (plane-of-the-sky) velocities are reported. Both flare locations and CME directions are expressed in terms in the Heliocentric Earth Equatorial (HEEQ) coordinate system.

\begin{figure}\center
\includegraphics[width=1\linewidth]{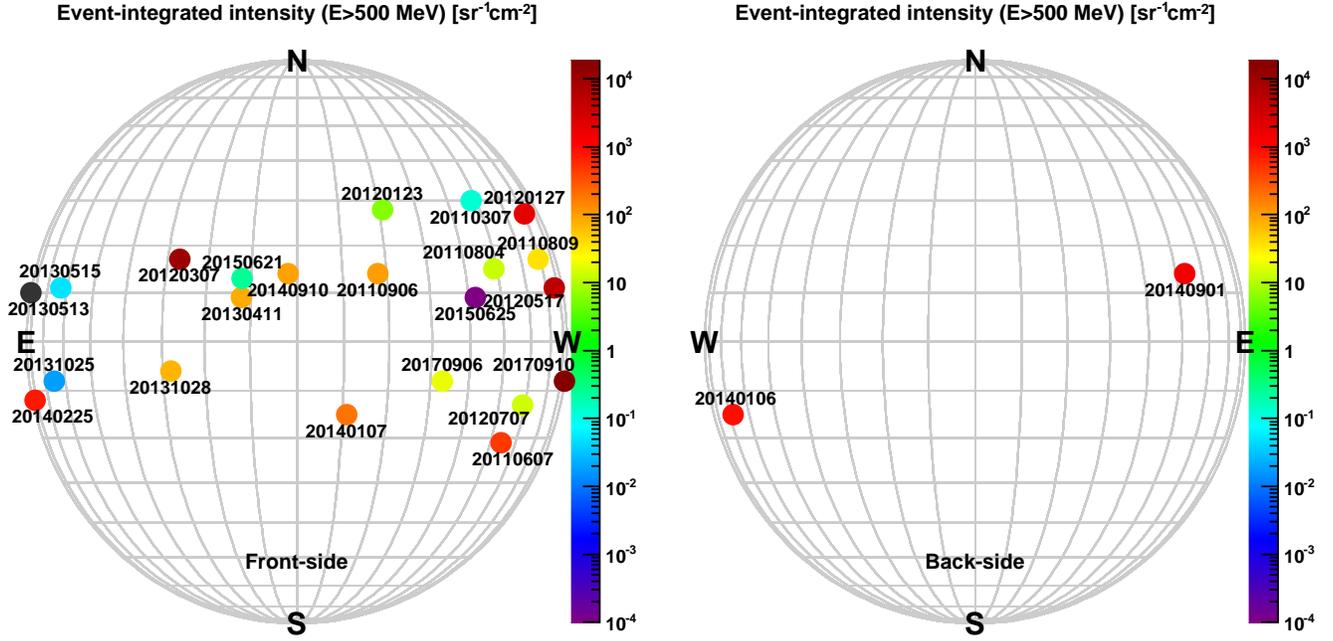}
\caption{Heliographic locations of parent flares for the SEP events listed in Table \ref{tab:sep-events}. Front-side and back-side events are displayed in the left and in the right panel, respectively. The color code indicates the SEP event-integrated intensity above 500 MeV; the one event with no signal above this threshold is shown in black.} 
\label{fig:helio_locations}
\end{figure}

All the events are associated with $\ge$M-class flares with hard X-ray emission extending above 100 keV \citep{ref:SHARE2018}, and with
full halo CMEs in the CDAW catalog.
In addition, they were linked to long-duration type-II and type-III radio bursts, indicating the presence of a shock and of open field lines, respectively. In particular, the measured type-II emission ranges from metric to decameter-hectometric wavelengths 
for most events \citep{ref:MITEVA2017}, suggesting that the shocks accelerating particles formed close to the Sun \citep{ref:GOPALSWAMY2017}. 
The heliographic locations of the flares associated with the SEP events listed in Table \ref{tab:sep-events} are shown in Figure \ref{fig:helio_locations}, for both front- and back-side events. The color code indicates the SEP event-integrated intensity above 500 MeV.
As expected based on magnetic connectivity considerations, most eruptions populate the western hemisphere; 
however, eruptions on the eastern hemisphere also contribute to the SEP flux detected near the Earth, in particular
two long-duration high-energy events from near the east limb were observed on 2014 February 25 and 2014 September 1.

\begin{figure}\center
\begin{tabular}{c}
\includegraphics[width=0.9\linewidth]{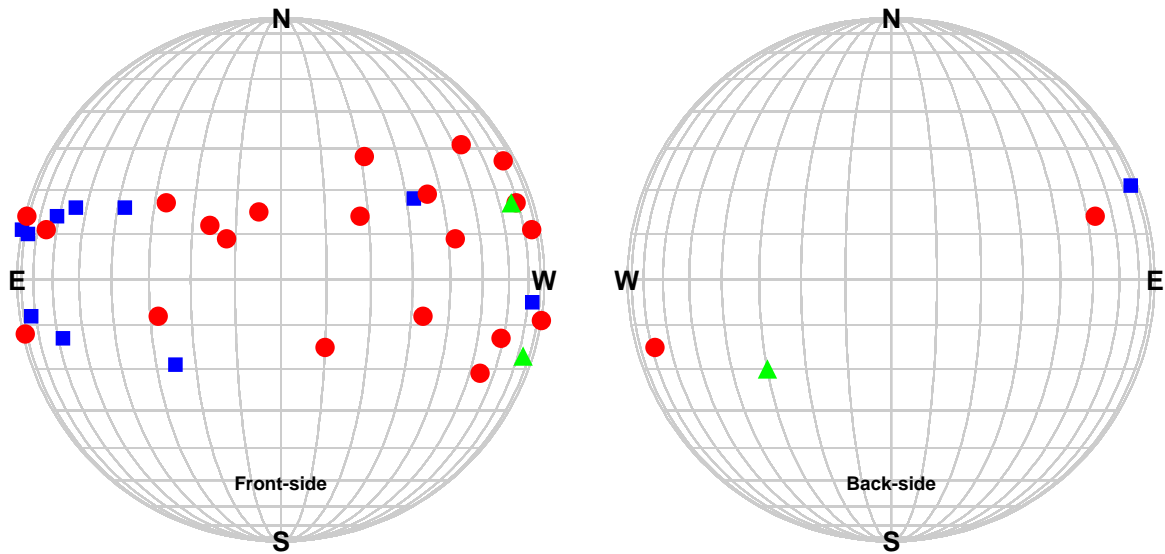}
\end{tabular}
\caption{Heliographic distribution of solar flares associated with LDGRFs detected by $\it{Fermi}$/LAT and SEPs measured by near-Earth spacecraft. The red circles denote the eruptions linked to both high-energy $\gamma$-ray and SEP events (see Table \ref{tab:sep-events}), while the blue squares correspond to eruptions with no registered SEP event (see Table \ref{tab:lat-only-events}); for comparison, the green triangles indicate three eruptions associated with SEP events with a statistically significant proton signal above 500 MeV but not linked to $>$100 MeV $\gamma$-ray emission.}
\label{fig:Heliographic_Fermi}
\end{figure}	

\begin{table}[!t]
\center
\footnotesize
\scriptsize
\setlength{\tabcolsep}{6pt}
\renewcommand{\arraystretch}{1.0}
\begin{tabular}{c|c|c|c|c|c|c|c|c|c|c}
& \multicolumn{3}{c|}{Flare} & \multicolumn{4}{c|}{CME} & \multicolumn{3}{c}{SEP signal}\\
\hline
No & Onset & Location & Class & 1$^{st}$ app. & Speed & Width & Direction & STB & Earth & STA\\
\hline
1 & 2011 06/02, 07:22 & S18E22 & C3.7 & 03/07, 08:12 & 1147 & 360 & S05E30 & $\lesssim$40 MeV & \nodata & \nodata\\
2 & 2011 09/07, 22:32 & N18W32 & X1.8 & 09/07, 23:06 & 792 & 360 & N28W40 & bkgr? & bkgr? & bkgr?\\
3\tablenotemark{$\star$} & 2011/09/24, 09:21 & N14E61 & X1.9 & 09/07, 09:48 & 1936\tablenotemark{$\natural$} & 145 & S09E50 & $>$60 MeV & bkgr? & \nodata \\
4 & 2012 03/05, 02:30 & N16E54 & X1.1 & 03/05, 04:00 & 1531 & 360 & N22E70 & $>$60 MeV & bkgr? & \nodata\\
5 & 2012 06/03, 17:48 & N15E38 & M3.3 & 06/03, 18:12 & 605\tablenotemark{$\natural$} & 180 & \nodata & $>$60 MeV & \nodata & \nodata\\
6\tablenotemark{$\star$} & 2012/10/23, 03:13 & S15E57 & X1.8 & \nodata & \nodata & \nodata & \nodata & \nodata & \nodata & \nodata \\
7 & 2012 11/27, 15:52 & N05W73 & M1.6 & \nodata & \nodata & \nodata & \nodata & \nodata & \nodata & \nodata \\
8 & 2013 05/13, 01:53 & N10E89 & X1.7 & 05/13, 02:00 & 1270 & 360 & N20E94 & $>$60 MeV & \nodata & \nodata\\
9 & 2013 05/14, 00:00 & N10E89 & X3.2 & 05/14, 01:25 & 2645 & 360 & N22E90 & bkgr? & bkgr? & \nodata\\
10 & 2013 10/11, 07:01 & N21E103 & M4.9 & 10/11, 07:24 & 1208 & 360 & S01E106 & $>$60 MeV & \nodata & \nodata\\
11 & 2013 10/25, 07:53 & S08E73 & X1.7 & 10/25, 08:12 & 599 & 360 & S02E67 & $>$60 MeV & \nodata & \nodata\\
\hline
\end{tabular}
\\
\footnotesize{(\textit{$\star$}) Impulsive $\gamma$-ray emission. (\textit{$\natural$}) Projected speed (partial halo CMEs).}
\renewcommand{\arraystretch}{1.2}
\caption{List of LDGRFs detected by $\it{Fermi}$/LAT \citep{ref:SHARE2018}, 
without a clearly observed associated SEP event at near-earth spacecraft.
The first column gives the event number. The next three columns report the flare onset (UT), location (deg) and class based on GOES soft X-ray data. Columns 5--8 indicate the associated CME first appearance time (UT), space speed (km s$^{-1}$) and angular width (deg) according to the CDAW catalog, and direction (deg) from DONKI. Columns 9--11 display the maximum SEP energy reported by STEREO-B, near-Earth spacecraft and STEREO-A, respectively; 
``bkgr?'' indicates the presence of a high background from a previous event may have obscured any SEP signal enhancement. The dots (...) indicate no data available (missing CME or SEP event).}
\label{tab:lat-only-events}
\end{table}

Figure \ref{fig:Heliographic_Fermi} displays the heliographic distribution of solar events accompanied by LDGRFs registered by $\it{Fermi}$/LAT. In particular, the red circles show the locations of eruptions associated with SEP events measured by near-Earth spacecraft (see Table \ref{tab:sep-events}), while the blue squares indicate the events for which no significant SEP signal was detected. 
The latter set, listed in Table \ref{tab:lat-only-events}, corresponds essentially to poorly connected events, concentrated in the eastern hemisphere.
In particular, columns 9--11 indicate the the energy range of SEP observations made by STEREO-A/B and near-Earth spacecraft.
The high background from a previous event (referred as ``bkgr?'') may have obscured any SEP signal in case of the eruptions on 2011 September 7 and 24, 2012 March 5 and 2013 May 14.
In addition, for five events -- 2011 June 2, 2012 June 3, 2013 May 13, 2013 October 11 and 2013 October 25 -- STEREO-B reported some moderate or large SEP enhancements.
However, for the two remaining events on 2012 October 23 and November 27 no SEP signal was observed by STEREO and near-Earth spacecraft; above all, these events
are peculiar because they were not linked to CMEs; \citet{ref:SHARE2018} reported an association with magnetic eruptions that may indicate failed CMEs. Nevertheless, such events suggest that the association with fast CMEs is not a necessary requirement for LDGRFs. 

\begin{figure}\center
\begin{tabular}{cc}
\includegraphics[width=0.5\linewidth]{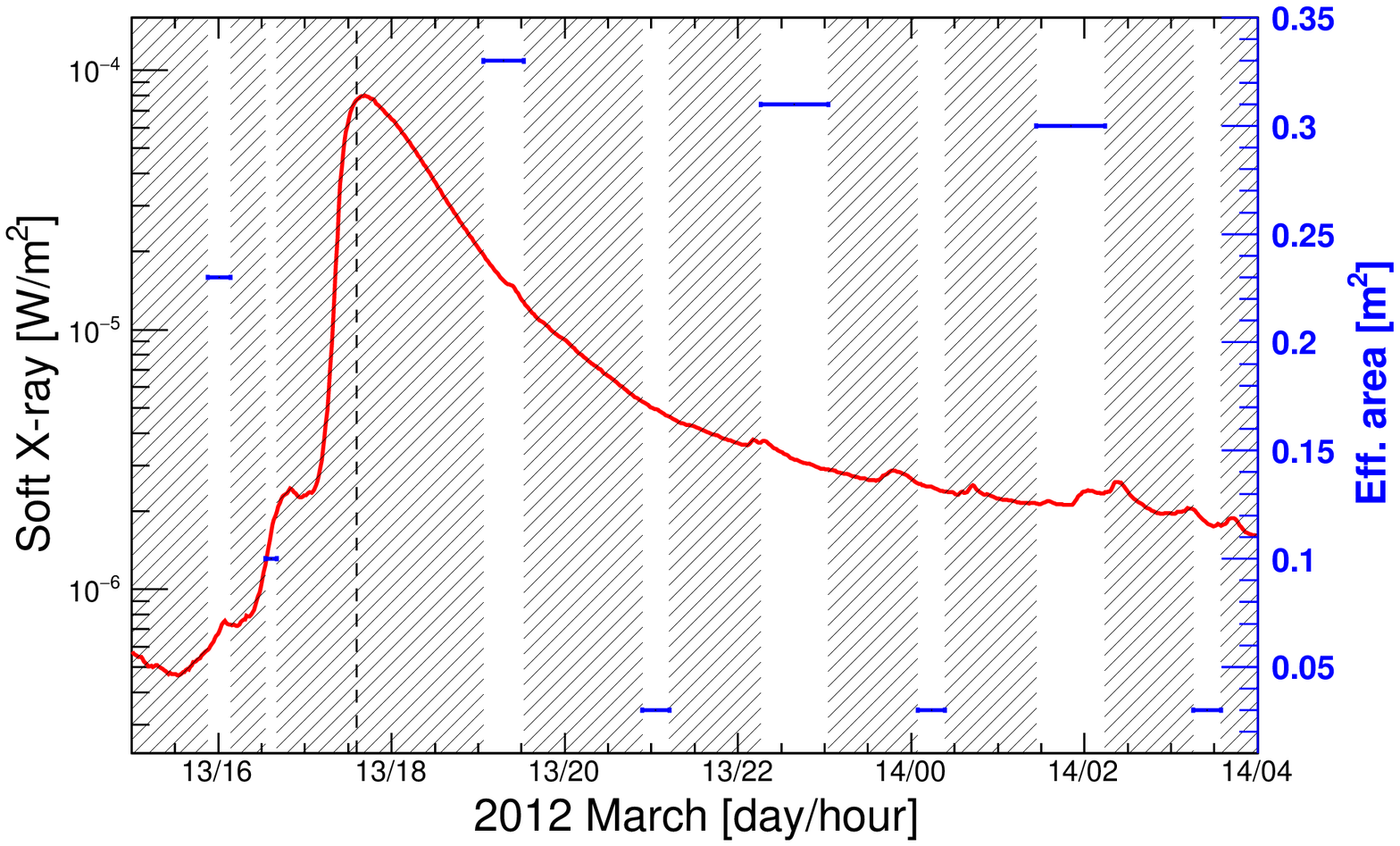} &
\includegraphics[width=0.5\linewidth]{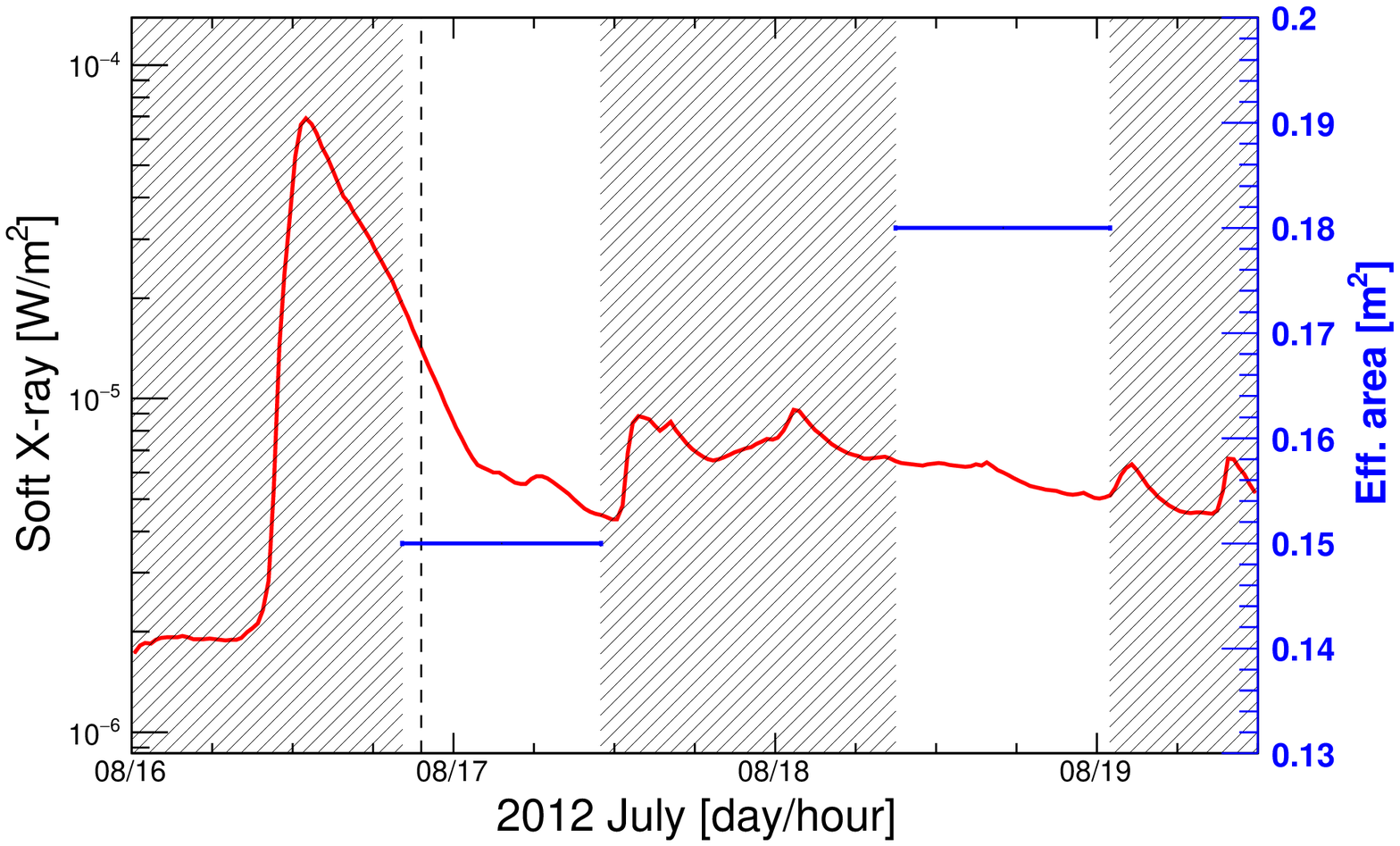}\\
\end{tabular}
\caption{Time profiles of soft X-ray emission during the 2012 March 13 and the 2012 July 8 eruptions. The vertical dashed lines refer to the related CME first appearance times in the LASCO C2 coronagraph. The horizontal blue lines indicate the average effective area of $\it{Fermi}$/LAT, while the shaded regions show the time intervals in which the Sun was occulted in the instrument field of view.} \label{fig:fermi_exposure}
\end{figure}

For comparison, the green triangles in Figure \ref{fig:Heliographic_Fermi} denote three eruptions without $\it{Fermi}$/LAT detections, linked to SEP events with a statistically significant proton signal with energies in excess of 500 MeV, the 2012 March 13, the 2012 July 8 and the 2015 October 29 events. 
The first two originated close to the western limb and were associated with an M7.9 flare and a full halo CME with a 1884 km s$^{-1}$ space speed, and with 
an M6.9 flare and a partial halo CME with a 1497 km s$^{-1}$ projected speed, respectively
\citep{ref:BRUNO2018}.
However, any firm conclusion about attendant $>$100 MeV $\gamma$-ray emission is precluded by the limited exposure of the LAT instrument, which can
monitor the Sun for 20--40 contiguous minutes every 1--2 hours depending on the precession of the orbit and the variation of the Sun latitude (\url{https://hesperia.gsfc.nasa.gov/fermi/lat/lat_solar_exposure_times.txt}). 
To illustrate this point, the time profiles of soft X-ray emission associated with the 2012 March 13 and the 2012 July 8 flares are shown in Figure \ref{fig:fermi_exposure}; the vertical dashed lines refer to the related CME first appearance times; the horizontal blue lines indicate the average effective area of $\it{Fermi}$/LAT, while the shaded regions show
the time intervals in which the Sun was occulted in the instrument field of view. Consequently, it can be speculated that $\it{Fermi}$/LAT may have missed the detection of relatively short-duration ($\lesssim$1 hr) $\gamma$-ray flares.
The third eruption producing a $>$500 MeV SEP event measured by near-Earth spacecraft without $\it{Fermi}$/LAT detection occurred on 2015 October 29; however, in this case it was located well behind the western limb (S20W150).
Finally, 
no LDGRF was detected
during the PAMELA SEP events with proton energies in excess of 300 MeV but
without a statistically appreciable signal above 500 MeV (not shown in Figure \ref{fig:Heliographic_Fermi}), such as the events on 2012 July 19, 2013 May 22 and 2014 April 18; these were linked to moderately bright X-ray flares (M7.7, M5.0 and M7.3 class, respectively) and, interestingly, fast (1631 km s$^{-1}$, 1491 km s$^{-1}$ and 1359 km s$^{-1}$ space speed, respectively) halo CMEs \citep{ref:BRUNO2018}.

\section{Deriving Total Proton Numbers from SEP Events}\label{Deriving Total Proton Numbers from SEP Events}
\subsection{Evaluation of the SEP spatial distribution}
Estimating the total proton number in space 
from near-Earth measurements of SEP fluxes
requires knowledge of the SEP spatial distribution at 1 AU and the mean free path during transport.
The longitudinal spread of SEPs has generally been attributed to the large spatial extent of the associated CME-driven shock 
\citep{ref:MASON1984,ref:CANE1988} with the earliest estimates inferred from single spacecraft measurements and comparisons with the associated active region \citep{ref:CANE1986,ref:REAMES1999,ref:VanHollebeke1975}. However, examples exist of events whose extent is wider than that suggested by the CME shock (e.g., \citet{ref:CLIVER1995,ref:CLIVER2005}) as well as examples of highly prompt intensity increases at widely separated spacecraft (e.g., the 2011 November 3 event \citep{ref:MEWALDT2013,ref:RICHARDSON2014}). 
Propagation across the nominal Parker spiral magnetic field is supposed to be enhanced by transport effects in interplanetary space, including cross-field diffusion and early-time non-diffusive propagation in turbulent fields along meandering field lines (see \citet{ref:DESAIGIACALONE2016,ref:Laitinen2018} and references therein).
However, the mechanisms that lead to rapid and efficient transport of SEPs within the heliosphere are still not well understood. 

Recently, several authors have investigated the SEP spatial distributions by taking advantage of multi-point observations 
\citep{ref:RICHARDSON2014,ref:RICHARDSON2017,ref:Lario2006,ref:Lario2013,ref:COHEN2017}. These studies show that the SEP peak or event-integrated intensity decreases with increasing longitudinal 
separation between the solar source and the footpoints of the IMF line connected to the observing spacecraft. With the limited number of observation points available, the longitudinal distribution cannot be determined directly, but is usually assumed to be Gaussian.
\citet{ref:RICHARDSON2014} obtained an average standard deviation $\sigma$ = 43 deg for protons between 14--25 MeV, consistent with a separate analysis by \citet{ref:Lario2013} for the peak intensity of protons between 25--53 MeV. 
\citet{ref:Lario2006} compared the longitudinal spread of peak and event-integrated intensities and found that their Gaussian standard deviations were not appreciably different. 
\citet{ref:COHEN2017} performed a fit of event-integrated intensities to periodic Gaussian distributions at two and three-spacecraft locations and determined an average standard deviation of 43 $\pm$ 1 deg that decreases with increasing energy. \citet{ref:RICHARDSON2017} 
bracket the most intense of the nearly thousand SEP events in their study with
a Gaussian standard deviation of $\sigma$ = 43 deg and suggest that this width is a good indicator of the upper limit of the intensity of 25 MeV proton events as a function of longitude of the solar event with respect to the observer.

In this study, the SEP longitudinal distribution is inferred by combining the event-integrated intensities measured by near-Earth spacecraft (PAMELA and GOES-13/15) and by STEREO-A/B. In case of STEREO, the 
SEP spectra are evaluated by using the data from
the Solar Electron and Proton Telescope (SEPT; 0.084--6.5 MeV, 10-min resolution), the Low Energy Telescope (LET; 4--12 MeV, 10-min resolution) and the High Energy Telescope (HET; 13.6--100 MeV, 15-min resolution), according to the procedure described in \citet{ref:BRUNO2019}; upper limits on the event-integrated intensities are computed by extrapolating the spectral fits to higher energies. 
Since STEREO measurements are limited to 100 MeV, using the intensities extrapolated to above 500 MeV would result in large uncertainties, producing a significant overestimate of the SEP longitudinal spread. Consequently, 
we evaluate upper limits for the $>$500 MeV event-integrated intensity distributions by examining the values obtained for energies higher than 80 MeV (PAMELA threshold).

Specifically, we use a periodic Gaussian function with the form:
\begin{equation}\label{eq:periodic_Gaussian}
G(\delta) = \frac{1}{3}exp\left(-\frac{\delta^{2}}{2\sigma^{2}}\right) + \frac{1}{3}exp\left(-\frac{\delta_{+}^{2}}{2\sigma^{2}}\right) + \frac{1}{3}exp\left(-\frac{\delta_{-}^{2}}{2\sigma^{2}}\right),
\end{equation}
with $\delta$ given by the great-circle or orthodromic distance from the peak of the SEP spatial distribution ($\beta_{sep}$, $\alpha_{sep}$):
\begin{equation}\label{eq:great-circle}
\delta = arccos\left[ sin(\alpha)sin(\alpha_{sep}) + cos(\alpha)cos(\alpha_{sep})cos(\beta-\beta_{sep})\right],
\end{equation}
where $\beta$ and $\alpha$ are the HEEQ longitude and latitude, 
$\sigma$ is the distribution standard deviation, and the terms associated with
\begin{equation}\label{eq:great-circle-pm}
\delta_{\pm} = arccos\left[ sin(\alpha)sin(\alpha_{sep}) + cos(\alpha)cos(\alpha_{sep})cos(\beta-\beta_{sep}\pm2\pi)\right]
\end{equation}
account for the possible contribution from particles propagating at angles $>$180 deg from the center of the distribution, ensuring that 
$G(\delta)$ = $G(\delta\pm2\pi)$
\citep{ref:COHEN2017}. The use of a periodic function is important especially for SEP events characterized by a broad longitudinal distribution.
In previous studies, the latitudinal magnetic connectivity to the SEP sources \citep{ref:DALLA2010,ref:GOPALSWAMYMAKELA2014}
was typically neglected compared to longitudinal variations, and the inclination of the propagation axis was ignored when fitting spatial distributions.
However, we note that, unless $\alpha_{sep}$ is null, the function describing the distribution projection on the X-Y plane 
is not Gaussian but includes a factor depending on $\alpha_{sep}$:
\begin{equation}\label{eq:GaussianProj}
G(\beta) = \frac{1}{3}exp\left\{ -\frac{arccos^{2}\left[cos(\alpha_{sep})cos(\beta-\beta_{sep})\right]}{2\sigma^{2}}\right\} + . . .,
\end{equation}
where 
the two terms centered at $\beta_{sep}\pm2\pi$ are omitted for brevity.
Consequently, fitting the particle intensities measured by spacecraft at 1 AU with a Gaussian function will result in a systematically larger standard deviation for higher latitude events; in particular, for the 2012 January 23 and 27 we estimate a $\sim$3 deg ($\sim$8\%) difference.
To a first approximation, $\alpha_{sep}$ is assumed to coincide with the latitudinal angle $\alpha_{flare}$ describing the parent flare location (see column 7 in Table \ref{tab:sep-events}).

Particle transport through the interplanetary magnetic field 
is accounted for
by computing 
the footpoints of the Parker spiral field lines crossed by the spacecraft, which are
mapped ballistically back to 30 $R_{s}$ and down to the photosphere based on measured plasma speed data
by using the 
Predictive Science web-tools (\url{http://www.predsci.com/stereo/spacecraft_mapping.php} and \url{http://www.predsci.com/hmi/spacecraft_mapping.php});
as results are available with an 1-hour resolution, the calculation is performed for the hour before the flare onset.
However, we do not use magnetic footpoints at the photosphere
for the estimate of $N_{SEP}$ because the results would be model-dependent (see, e.g., \citet{ref:Lario2017}). 
Consistent with \citet{ref:COHEN2017}, the differences between footpoints obtained
at 30 $R_{s}$ and at the photosphere are of the order of $\sim$10 deg, which can be assumed as an indication of the associated uncertainty. 
On the other hand, significantly larger deviations are found for the 2014 February 25 event, for which the derived STEREO-A/B footpoints at the photosphere are very close (see Figure 18 in \citet{ref:COHEN2017}), in contrast to the large discrepancy in terms of detected particle intensities. 
Because the footpoints of the field lines connecting the spacecraft generally do not lie in the X-Y plane, and 
in order to apply Equation \ref{eq:GaussianProj} to the 
measured intensities, we apply a correction factor to account for the spacecraft footpoint latitude $\alpha_{sc}$:
\begin{equation}\label{footlat_corr}
K_{\alpha} = exp\left( \frac{ \delta_{sc}^{2}-\delta_{sc,0}^{2}}{2\sigma^{2}} \right),
\end{equation}
where $\delta_{sc}$ and $\delta_{sc,0}$ are the great-circle distances from the peak of the SEP spatial distribution evaluated with respect to the spacecraft footpoints
and to their projection on the X-Y plane ($\alpha_{sc}$=0), respectively.
Since $\sigma$ is unknown a priori, 
we perform an iterative procedure until fit results become stable.
We note that, even though $\alpha_{sc}$ is typically a few degrees, the differences in terms of great-circle distances are much larger, and thus 
the associated correction can be significant. 
Finally, the small discrepancies ($<$5\%) regarding radial distances are neglected and all the spacecraft are assumed to be located exactly at 1 AU.

\subsection{Estimate of transport effects in interplanetary space}
The transport of SEPs in the interplanetary medium is governed not only by the large scale magnetic field geometry but also by small-scale scattering from turbulence.
One consequence is that particles may pass the distance of the observing spacecraft several times
\citep{ref:ZANK2006}, requiring a correction of the flux and in turn the local number density. \citet{ref:ZANK2006} modeled the number of crossings at 1 AU, $N_{cross}(E)$, for SEPs accelerated by both weak and strong shocks and found a $\overline{N}_{cross}$$\sim$2--4 mean value above 100 MeV but, on occasion, high energies particles experience multiple crossings as high as 15. \citet{ref:Chollet2010} 
used numerical simulations of particle transport to 
determine $\overline{N}_{cross}$$\sim$6--7 for 100 MeV.

Since SEP intensity and anisotropy distributions depend
on the scattering mean free path $\lambda$ parallel to the magnetic field \citep{ref:Palmer1982}, we estimate
$\overline{N}_{cross}$ by means of simulations of relativistic proton propagation 
under
a variety of scattering 
conditions. We consider two separate test particle simulation codes, the first (Simulation Code 1, SC1) presented by \citet{ref:Chollet2010} and the second (Simulation Code 2, SC2) by \citet{ref:Battarbee2018}.
For both codes, we assumed an impulsive injection of mono-energetic isotropic particles at 0.1 AU and followed them
for ten days, and include magnetic focusing and scattering off an unspecified plasma turbulence field. 
Particle crossings over the entire 1 AU sphere are added together and averaged over the mono-energetic population considered.

Using SC1, we let the turbulence take one of two forms, either uniform from the launch radius to 1 AU (Turbulence Model 1),
or varying in proportion to the gyro-cyclotron radius $r_{g}$ (Turbulence Model 2). 
The resulting transport calculations predict the time-dependent development and decay of the intensity at 1 AU. The mean numbers
of crossings as a function of energy are reported in Table \ref{tab:transport_SC1}, for both turbulence models. A flat heliospheric current sheet (HCS) is assumed, and results are valid for a positive solar magnetic field polarity in the northern hemisphere. 
It should be noted that $\overline{N}_{cross}$ distributions are very broad, as the associated RMS values (not shown in the Table) are of the same magnitude as the mean values.

The average number of 1 AU crossings was also calculated by means of SC2 \citep{ref:Marsh2013,ref:Battarbee2018}, a code which can include the effects of a wavy HCS.
We use $\lambda$=const (Turbulence Model 1) for these simulations. 
Table \ref{tab:transport_SC2} shows how the number of crossings varies for different configurations of the HCS, including no HCS, flat or
wavy HCS. Using the standard galactic cosmic ray definition, A$^{+}$ refers to a situation in which the
polarity of the interplanetary magnetic field is positive (outward) in the northern hemisphere and negative (inward) in
the southern hemisphere, and the opposite for A$^{-}$. 
The number of crossings for E=1000 MeV, $\lambda_{0}$ =0.1 AU, flat HCS and A$^{+}$ compares well with that of SC1, 
as shown in Table \ref{tab:transport_SC1} (corresponding to a run with flat HCS and A$^{+}$), showing that the two codes are in good agreement.
Table \ref{tab:transport_SC2} shows that at particle energies of 1 GeV
there is a large difference in the number of crossings for the A$^{+}$ and A$^{-}$ cases. The difference is
similar both for the case of a flat and wavy HCS. The cause of this difference is the particle drift along the HCS,
which for the A$^{+}$ case helps to move protons outward from the inner heliosphere faster. 
A full simulation of the 2012 May 17 event, with the initial proton distribution given by a power law,
shows good agreement between the SC2 and PAMELA intensity time profiles when $\lambda_{0}$ =0.3 AU (Dalla et al. 2019, in preparation).

The solar polarity reversal during cycle 24 was unusually complex. In fact, based on the measurements of
Wilcox Solar Observatory, the northern polar field changed polarity in 2012 June, while the southern
polar field reversed in 2013 July; however, additional analyses of solar data suggest that the field reversal
was completed in 2014--2015 (see, e.g., \citet{ref:JANARDHAN2018} and references therein).

\begin{table}
\scriptsize
\setlength{\tabcolsep}{6pt}
\renewcommand{\arraystretch}{1.0}
\center
\begin{tabular}{c|c|c|c|c}
Energy & \multicolumn{2}{c|}{Turbulence Model 1} & \multicolumn{2}{c}{Turbulence Model 2} \\
(MeV) & \scriptsize{$\lambda=const$, $\lambda_{0}=0.1$ AU} & \scriptsize{$\lambda=const$, $\lambda_{0}=0.5$ AU} & \scriptsize{$\lambda\propto r_{g}$, $\lambda_{0}=0.1$ AU} & \scriptsize{$\lambda\propto r_{g}$, $\lambda_{0}=0.5$ AU}\\
\hline
500 & 18.4 & 7.0 & 14.5 & 4.7 \\
1000 & 16.5 & 5.9 & 11.8 & 4.1 \\
2000 & 17.2 & 5.0 & 10.6 & 3.6 \\
\hline
\end{tabular}
\caption{Average number of 1 AU crossings at sample proton energies, as predicted by SC1 \citep{ref:Chollet2010} for different turbulence models.}
\label{tab:transport_SC1}
\end{table}

\begin{table}
\center
\scriptsize
\setlength{\tabcolsep}{6pt}
\renewcommand{\arraystretch}{1.0}
\begin{tabular}{c|c|c|c|c}
Energy & $\lambda_{0}$ & HCS & Magnetic & $ \overline{N}_{cross} $ \\
(MeV) & (AU) & configuration & polarity & \\
\hline
1000 & 0.1 & no HCS & + both poles & 31 \\
1000 & 0.1 & no HCS & -- both poles & 28 \\
\hline
500 & 0.1 & flat HCS & A$^{+}$ & 19 \\
500 & 0.1 & flat HCS & A$^{-}$ & 29 \\
500 & 0.5 & flat HCS & A$^{+}$ & 8 \\
500 & 0.5 & flat HCS & A$^{-}$ & 11 \\
1000 & 0.1 & flat HCS & A$^{+}$ & 15 \\
1000 & 0.1 & flat HCS & A$^{-}$ & 28 \\
\hline
500 & 0.1 & wavy HCS & A$^{+}$ & 21 \\
500 & 0.1 & wavy HCS & A$^{-}$ & 30 \\
500 & 0.5 & wavy HCS & A$^{+}$ & 8 \\
500 & 0.5 & wavy HCS & A$^{-}$ & 11 \\
1000 & 0.1 & wavy HCS & A$^{+}$ & 17 \\
1000 & 0.1 & wavy HCS & A$^{-}$ & 29 \\
1000 & 0.5 & wavy HCS & A$^{+}$ & 7 \\
1000 & 0.5 & wavy HCS & A$^{-}$ & 11 \\
\hline
\end{tabular}
\caption{Average number of 1 AU crossings at sample proton energies, as predicted by SC2 \citep{ref:Battarbee2018} for different types of HCS configurations,
mean free path values and solar magnetic polarities. A$^{+}$ corresponds to positive polarity north pole and negative polarity south pole, while the opposite holds for A$^{-}$; ``+/-- both poles'' denote a positive/negative polarity for both poles.}
\label{tab:transport_SC2}
\end{table}

\begin{figure}\center
\includegraphics[width=0.8\linewidth]{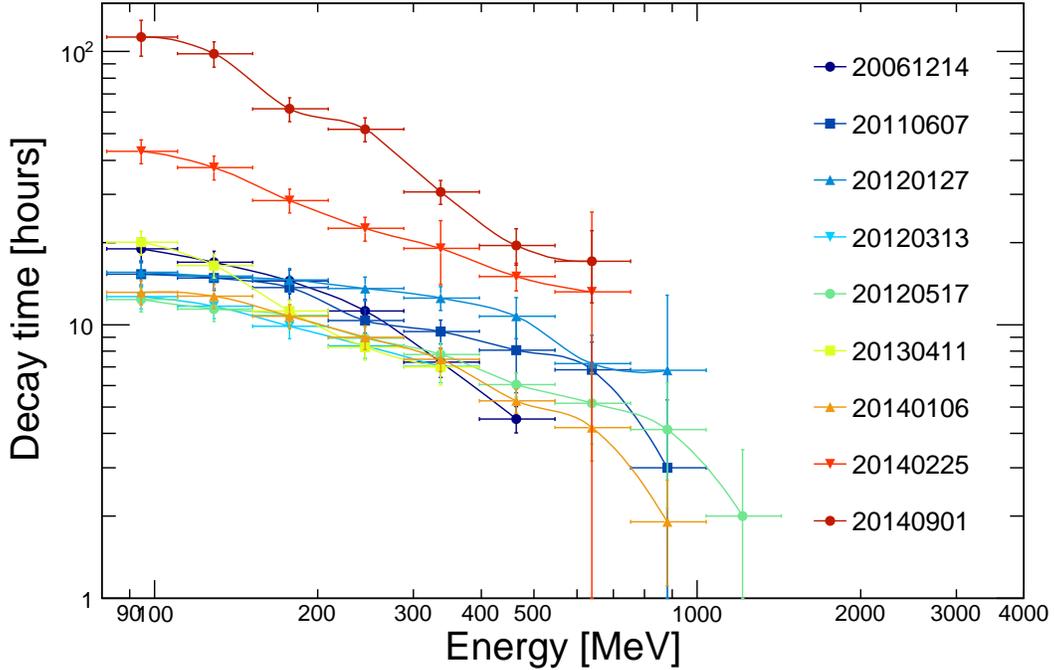}
\caption{Decay times of proton intensities as a function of energy, for sample SEP events measured by PAMELA. The curves are to guide the eye.} \label{fig:decay_times}
\end{figure}

\begin{table}[!h]
\center
\scriptsize
\setlength{\tabcolsep}{6pt}
\renewcommand{\arraystretch}{1.0}
\begin{tabular}{c|c|c|c|c}
& \multicolumn{2}{c|}{Turbulence Model 1} & \multicolumn{2}{c}{Turbulence Model 2} \\
Energy & \scriptsize{$\lambda=const$, $\lambda_{0}=0.1$ AU} & \scriptsize{$\lambda=const$, $\lambda_{0}=0.5$ AU} & \scriptsize{$\lambda\propto r_{g}$, $\lambda_{0}=0.1$ AU} & \scriptsize{$\lambda\propto r_{g}$, $\lambda_{0}=0.5$ AU} \\ 
(MeV) & $\tau$ (hours) & $\tau$ (hours) & $\tau$ (hours) & $\tau$ (hours) \\
\hline
100 & $>$15.5 & 2.8 & 10.6 & 1.5 \\
200 & 13.1 & 2.1 & 8.7 & 1.3 \\
500 & 6.5 & 1.5 & 5.7 & 1.1 \\
1000 & 4.8 & 1.4 & 3.7 & 1.0 \\
2000 & 2.6 & 1.1 & 2.6 & 1.0 \\
\hline
\end{tabular}
\caption{Decay times for 1 AU intensity time profiles at sample proton energies predicted by SC1 \citep{ref:Chollet2010} for different turbulence models.}
\label{tab:decaytime_SC1}
\end{table} 

Figure \ref{fig:decay_times} shows decay times deduced from the time-intensity profiles for the SEPs observed by PAMELA. With the exception of the two eastern events on 2014 February 25 and 2014 September 1, for which we expect significant cross field diffusion, the decay times all show similar energy dependence. 
Table \ref{tab:decaytime_SC1} gives values of the decay times from SC1, for the two turbulence models considered. As expected from diffusive transport theory, a shorter mean free path results in a longer decay time. The SC1 values that best match the PAMELA observations are those for Turbulence Model 1 ($\lambda$ = constant), with a mean-free path at 1 AU of $\lambda_{0}$ = 0.1 AU.
Simulation results are in a better agreement with the predictions of Model 1 for $\lambda$ = constant. In particular, the analysis of the 2012 May 17 event suggests a value of $\sim$0.3 AU for the mean free path.
For the purpose of this work, because we compute upper limits on the number of SEPs at 1 AU (see below),
we conservatively assume $\lambda_{0}$ = 0.5 AU and we use the numbers of crossings estimated at 500 MeV with the wavy HCS configuration and A$^{-}$/A$^{+}$ magnetic polarity for SEP events occurring before/after 2012 June, respectively (see Table \ref{tab:transport_SC2}).
Finally, we note that model calculations do not include possible effects related to local solar wind structures, including magnetic mirroring from nearby reflecting boundaries (see, e.g., \cite{ref:TAN2009}).

\subsection{Assessment of the number of SEP protons at 1 AU}
After estimating the SEP spatial distribution at 1 AU and the mean free path during transport, we can derive
the total number of solar protons above 500 MeV 
using the following relation:
\begin{align}\label{eq:Nsep1}
N_{SEP} & = \overline{N}_{cross}^{-1} \int_{4\pi}d\Omega \int_{S}dS\left( \boldsymbol{J \cdot n} \right) \\
& = \overline{N}_{cross}^{-1} \int_{4\pi}d\Omega \int_{S}dS \hspace{0.1cm}cos(\theta) \hspace{0.1cm}J(\Omega,S),
\end{align}
where 
$S$ is the heliocentric spherical surface with radius $R_{o}$ = 1 AU, 
$d\Omega$ = $d\phi d\theta sin(\theta)$ is the solid angle element 
with polar angles $\phi$ and $\theta$ defining particle velocity direction in the reference frame centered at a point on the sphere,
 and $J=J(\Omega,S)$ is the event-integrated intensity for energies $>$500 MeV; 
the dot-product accounts for the fact that the flux at an angle $\theta$ with respect to the local normal to the sphere surface ($n$ is the unit vector) is proportional to $cos(\theta)$. Consequently,
for an isotropic flux, $J(\Omega,S)$ = $J(S)$ (independent on $\Omega$), 
the integration over the 4$\pi$ detection solid angle subtended at a point on the sphere gives:
\begin{equation}\label{eq:Ang_integration}
 \int_{4\pi}d\Omega \hspace{0.1cm}cos(\theta) \hspace{0.1cm}J(\Omega,S) = \int_{0}^{2\pi}d\phi \int_{0}^{\pi}d\theta \hspace{0.1cm}sin(\theta) cos(\theta)\hspace{0.1cm} J(\phi,\theta,S) = 2\pi J(S).
\end{equation}
Integrating Equation \ref{eq:Ang_integration} over the whole spherical surface and
assuming a Gaussian spatial distribution 
$J(S)$ = $J_{max}G(\delta)$, where the amplitude
$J_{max}$ is 
the maximum particle intensity at 1 AU (corresponding to the SEP propagation axis) and $G(\delta)$ is given by Equation \ref{eq:periodic_Gaussian}, 
Equation \ref{eq:Nsep1} reduces to:
\begin{align}
N_{SEP} & = 2\pi \hspace{0.1cm} \overline{N}_{cross}^{-1} \hspace{0.1cm} J_{max} \hspace{0.1cm} \int_{S}dS \hspace{0.1cm} G(\delta)\\
 & = 2\pi \hspace{0.1cm} \overline{N}_{cross}^{-1} \hspace{0.1cm} J_{E} \hspace{0.1cm} S_{J} \hspace{0.1cm} C_{spa},
\end{align}
where $J_{E}$ is the $>$500 MeV event-integrated 
intensity 
measured by PAMELA and $S_{J}$ is the spherical area weighted by the particle spatial distribution:
\begin{equation}\label{eq:Warea}
S_{J} = 2\pi R_{o}^{2} \int_{0}^{\pi}d\delta \hspace{0.1cm} sin\delta \hspace{0.1cm} G(\delta),
\end{equation}
where $\delta$ is the great-circle distance with respect to the peak of the SEP spatial distribution
(Equation \ref{eq:great-circle}).
The Gaussian standard deviation is derived by fitting the longitudinal distribution given by $>$80 MeV particle intensities measured by STEREO and PAMELA by means of the function given by Equation \ref{eq:GaussianProj}. The integration of \ref{eq:Warea} is then performed by means of numerical techniques.
Finally, $C_{spa}$ accounts for the fact that, in general, PAMELA's observations 
are not made on interplanetary magnetic field lines that connect with the peak of the particle distribution:
\begin{equation}\label{eq:SpatialCorr}
C_{spa} = exp\left( \frac{ \delta_{pam}^{2} }{2\sigma^{2}}\right),
\end{equation}
where $\delta_{pam}$ is the great-circle distance between PAMELA's magnetic footpoints and the peak of the SEP spatial distribution, so that:
\begin{equation}
J_{max} = J_{E} \hspace{0.1cm} C_{spa}.
\end{equation}
As a final remark, we note that the mathematical formulation described in this Section 
provides significantly more robust basis to our calculation,
representing a major improvement with respect to the simpler approaches adopted in previous studies.

\begin{figure}\center
\includegraphics[width=0.8\linewidth]{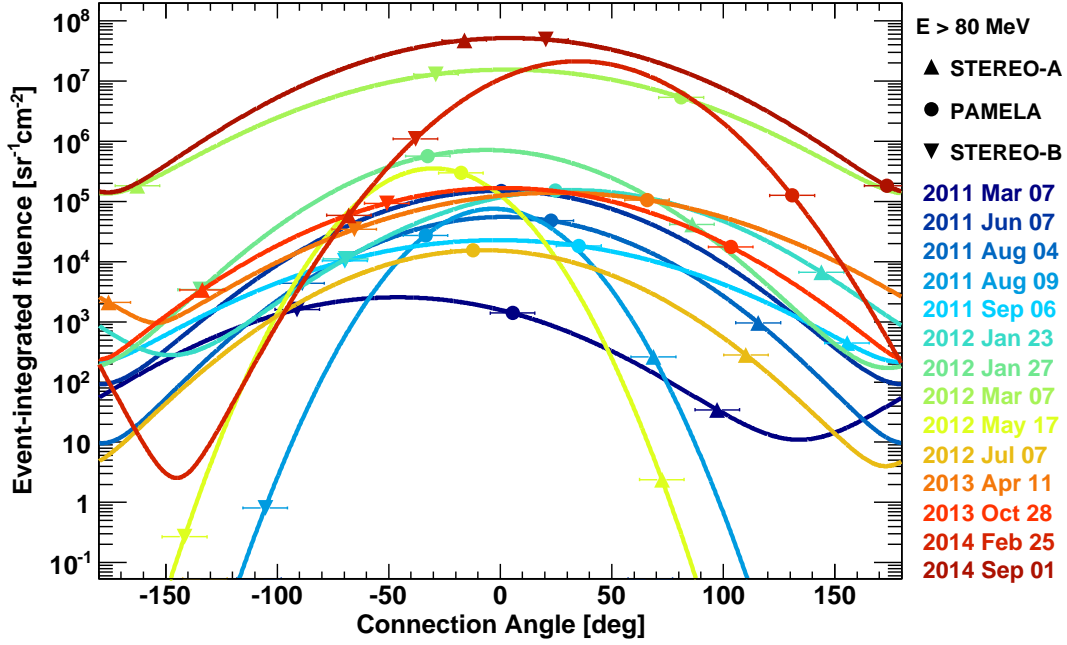}
\caption{Longitudinal extent of SEP events determined from the fits (Equation \ref{eq:GaussianProj}) of the event-integrated intensities ($>$80 MeV) measured by PAMELA and STEREO-A/B, as a function of the longitudinal difference (connection angle) between the spacecraft magnetic footpoints at 30 $R_{s}$ and the location of the parent flare.} \label{fig:Gaussian_fits}
\end{figure}

\section{Results}\label{Results}
\subsection{SEP spatial distributions}
The numbers of protons $N_{SEP}$ and $N_{LDGRF}$ were derived 
for the SEP events in common with the ``light bucket'' analysis by \citet{ref:SHARE2018} (see Table \ref{tab:sep-events})
with the exception of: the 2013 May 13 event, for which no appreciable $>$500 MeV SEP signal was measured by near-Earth spacecraft; 
the 2013 May 15 eruption, due to the high background from a previous event that may have obscured any clear SEP 
enhancement in the measurements of STEREO-B, the best-connected spacecraft; 
and the 2015 June 21 event, occurring after the loss of communications with STEREO-B in 2014 October.
The selected sample consists of fourteen events mostly located in the visible western hemisphere, but it also includes poorly connected events such as the 2014 February 25 eastern limb and the 2014 September 1 backside events.
The PAMELA and STEREO-A/B event-integrated intensities are displayed in Figure \ref{fig:Gaussian_fits}
as a function of the longitudinal difference $\Delta\beta$ = $\beta_{sc}$ -- $\beta_{sep}$ (connection angle) between the spacecraft magnetic footpoints at 30 $R_{s}$ and the location of the parent flare; thus positive (negative) angles correspond to footpoints west (east) of the flare.

SEP distributions are influenced by several factors. For the 2011 June 7 event the intensities measured by STEREO-A/B are dominated by the background from a previous event. Here, we assume a standard deviation of 40.4 deg which, as discussed below, represents the mean $\sigma$ 
that characterizes the connection-angle dependence of the event-integrated $>$80 MeV intensity spectra measured by PAMELA. The same standard deviation is assumed for the 2012 July 7 event, for which no significant SEP signal was measured by STEREO-B; the integration interval used for the intensities measured by PAMELA and STEREO-A is limited by the onset of a following SEP event on July 8, resulting in intensities that are underestimated. 
Similarly, the time integration over the 2012 January 23 event is limited by the onset of the January 27 event, and the intensities measured during the latter include a component from the previous event.
In the case of the 2014 September 1 event, large gaps present in STEREO-A data preclude constructing event-integrated intensities; based on the comparison of the respective time profiles, the SEP flux is assumed to be equal to the one measured by STEREO-B. The time-integrated intensities measured during the 2011 March 7 event include particles injected at three different eruptions. In particular, those registered by STEREO-B comprise a significant component from the previous well-connected eruption, originating few hours before. This translates to a longitudinal distribution that is narrower with a peak closer to the Earth's magnetic footpoints. Similarly, SEP intensities measured by STEREO-B and, to a lesser extent, also by STEREO-A during the 2013 October 28 event, include a contribution from another eruption occurring three days before, resulting in an overestimate of the total number of protons $N_{SEP}$ at 1 AU.

\begin{figure}
\center
\setlength{\tabcolsep}{0pt}
\includegraphics[width=1\linewidth]{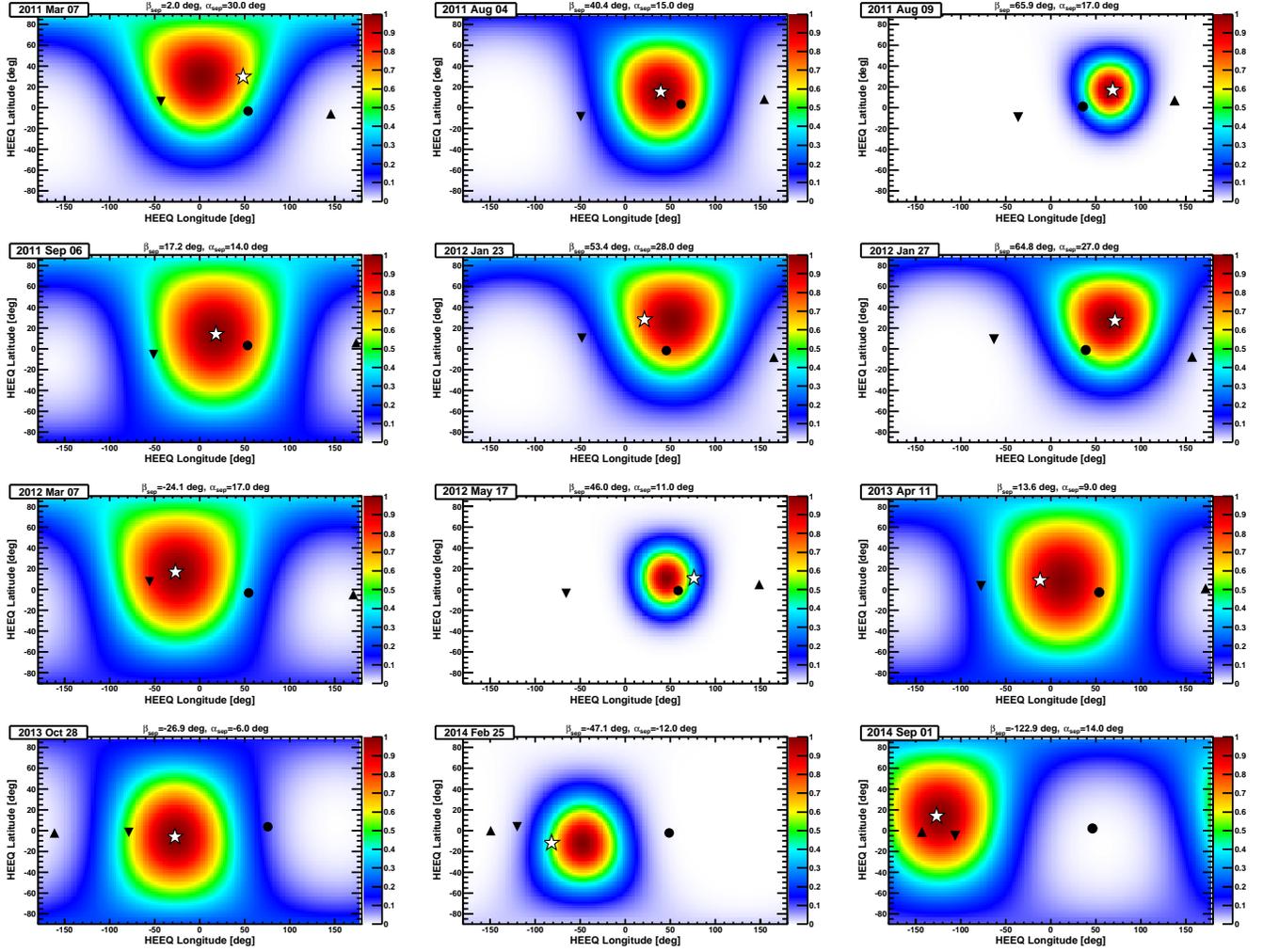} 
\caption{Normalized SEP spatial distributions in HEEQ coordinates $G(\delta)$ = $G(\beta,\alpha)$ (Equation \ref{eq:periodic_Gaussian}) inferred by the $>$80 MeV event-integrated intensity measurements. The stars show the parent flare locations, the full circles indicate PAMELA's magnetic footpoints, while upward- and downward-pointing triangles refer to those of STEREO-A and -B, respectively; only SEP events with three spacecraft measurements are displayed. The direction of the peak of the SEP spatial distribution is reported at the top of the panels.} \label{fig:maps}
\end{figure}

Some studies reported that the centers of the derived SEP distributions tend to lie west of the flare location \citep{ref:Lario2006,ref:Lario2013,ref:COHEN2017},
although this trend was not confirmed by other analyses (e.g., \citet{ref:RICHARDSON2014}).
\citet{ref:Lario2014} suggested that this displacement would be likely due to the fact that the maximum peak intensity is observed at some helioradial distance for which the eruption occurs eastward with respect to the footpoints of the nominal field line connecting the spacecraft with the Sun.
The effect is expected to be larger at lower energies as particles escape from the acceleration region at higher heights \citep{ref:COHEN2017}.
The SEP event sample analyzed in this work is characterized by significant variability, 
with eight (six) event distributions centered at positive (negative) connection angles $\Delta\beta$, 
resulting in a $\sim$0.7 deg mean angle between the flare and the peak of the SEP distribution, and a $\sim$22 deg RMS value.
In addition,
effects related to the multiple injections discussed above are also likely to contribute.
A majority of the SEP events with intensities reaching 1 GeV are confined to a narrow swath of longitudes, suggesting that the longitudinal spread at higher energies is even narrower than at 80 MeV, consistent with the energy dependence reported by \citet{ref:COHEN2017} below 10 MeV. Thus, the longitudinal distribution inferred from $>$80 MeV protons is an upper bound to the longitudinal extent at higher energies. 
This upper limit is also justified by the fact that we assume the same angular distribution for the latitudinal extent as the longitudinal dependence (i.e. $\sigma=\sigma_{\alpha}=\sigma_{\beta}$).
The normalized SEP spatial distributions $G(\delta)$ = $G(\beta,\alpha)$ (Equation \ref{eq:periodic_Gaussian}) are shown in Figure \ref{fig:maps} as a function of HEEQ coordinates. The stars show the parent flare locations, the circles indicate PAMELA's magnetic footpoints, while upward- and downward-pointing triangles show those of STEREO-A and -B, respectively. Only SEP events with three spacecraft measurements are displayed. The angles defining the direction of the peak of the SEP spatial distribution are reported at the top of the panels.

\begin{sidewaystable}[!t]
\centering
\footnotesize\scriptsize
\setlength{\tabcolsep}{2.5pt}
\renewcommand{\arraystretch}{1.0}
\begin{tabular}{c|c|cccc|ccc|cccc|cc|c|c|c|c|c}
\hline
 \multicolumn{2}{c|}{SEP event} & \multicolumn{4}{c|}{STEREO-B} & \multicolumn{3}{c|}{PAMELA} & \multicolumn{4}{c|}{STEREO-A} & \multicolumn{2}{c|}{Fit param.} & \multicolumn{1}{c|}{S$_{J}$} & \multicolumn{1}{c|}{$C_{spa}$} & \multicolumn{1}{c|}{$\overline{N}_{cross}$} & \multicolumn{1}{c|}{$N_{SEP}$} & \multicolumn{1}{c}{$N_{LDGRF}$}\\
No. & Date & Loc. & $\Delta\beta$ & $\Delta\alpha$ & Intensity & $\Delta\beta$ & $\Delta\alpha$ & Intensity & Loc. & $\Delta\beta$ & $\Delta\alpha$ & Intensity & $\beta_{sep}$ & $\sigma$ & & & & & \\
\hline
1 & 2011 Mar 07 & -93.9 & -91.0 & -23.8 & 1.6$\times$10$^{3\sharp}$ & 5.6 & -33.0 & 1.4$\times$10$^{3\dagger}$ & 88.7 & 97.3 & -36.1 & 3.5$\times$10$^{1}$ & 2.0 & 44.5 & 6.97$\times$10$^{26}$ & 2.42 & 11 & 1.18$\times$10$^{26}$ & (1.51$\pm$0.44)$\times$10$^{29}$ \\
2 & 2011 Jun 07 & -91.9 & -108.2 & 13.8 & bkgr & 0.5 & 21.0 & 1.5$\times$10$^{5}$ & 96.0 & 67.4 & 28.2 & bkgr & 54.5 & 41.0 & 6.08$\times$10$^{26}$ & 1.14 & 11 & 1.93$\times$10$^{29}$ & (2.20$\pm$0.80)$\times$10$^{28}$ \\
3 & 2011 Aug 04 & -91.9 & -88.8 & -23.5 & 4.4$\times$10$^{3}$ & 22.9 & -11.9 & 4.8$\times$10$^{4}$ & 101.3 & 115.7 & -6.8 & 9.6$\times$10$^{2}$ & 40.4 & 39.5 & 5.72$\times$10$^{26}$ & 1.21 & 11 & 5.04$\times$10$^{27}$ & (2.52$\pm$0.63)$\times$10$^{28}$ \\
4 & 2011 Aug 09 & -92.0 & -105.4 & -26.2 & 8.1$\times$10$^{-1}$ & -33.5 & -15.8 & 2.7$\times$10$^{4}$ & 101.8 & 68.7 & -10.1 & 2.6$\times$10$^{2}$ & 65.9 & 21.0 & 1.80$\times$10$^{26}$ & 3.69 & 11 & 1.41$\times$10$^{28}$ & (5.60$\pm$1.40)$\times$10$^{27}$ \\
5 & 2011 Sep 06 & -94.0 & -69.7 & -19.7 & 1.0$\times$10$^{4}$ & 35.2 & -11.1 & 1.8$\times$10$^{4\sharp}$ & 103.9 & 155.5 & -8.3 & 4.4$\times$10$^{2\sharp}$ & 17.2 & 54.2 & 9.42$\times$10$^{26}$ & 1.27 & 11 & 7.09$\times$10$^{28}$ & (5.06$\pm$0.92)$\times$10$^{28}$ \\
6 & 2012 Jan 23 & -113.0 & -69.3 & -17.7 & 1.1$\times$10$^{4\ast}$ & 24.7 & -29.6 & 1.5$\times$10$^{5\ast}$ & 108.8 & 144.0 & -36.1 & 6.7$\times$10$^{3\ast}$ & 53.4 & 42.0 & 6.35$\times$10$^{26}$ & 1.30 & 11 & 2.89$\times$10$^{27}$ & (6.60$\pm$1.32)$\times$10$^{28}$ \\
7 & 2012 Jan 27 & -113.6 & -134.5 & -17.9 & 3.4$\times$10$^{3\sharp}$ & -32.5 & -28.2 & 5.6$\times$10$^{5\sharp}$ & 108.9 & 86.0 & -34.7 & 4.2$\times$10$^{4\sharp}$ & 64.8 & 36.9 & 5.09$\times$10$^{26}$ & 1.70 & 11 & 9.74$\times$10$^{29}$ & (1.87$\pm$1.10)$\times$10$^{28}$ \\
8 & 2012 Mar 07 & -116.7 & -28.9 & -9.4 & 1.3$\times$10$^{7}$ & 81.1 & -20.0 & 5.3$\times$10$^{6}$ & 110.6 & -162.7 & -21.6 & 1.8$\times$10$^{5}$ & -24.1 & 52.6 & 9.03$\times$10$^{26}$ & 3.14 & 11 & 1.75$\times$10$^{31}$ & (3.81$\pm$0.68)$\times$10$^{30}$ \\
9 & 2012 May 17 & -116.8 & -141.6 & -14.7 & 2.7$\times$10$^{-1}$ & -17.6 & -12.2 & 3.0$\times$10$^{5}$ & 116.0 & 72.5 & -5.8 & 2.4$\times$10$^{0}$ & 46.0 & 20.8 & 1.78$\times$10$^{26}$ & 1.41 & 11 & 7.56$\times$10$^{29}$ & (2.10$\pm$1.40)$\times$10$^{27}$ \\
10 & 2012 Jul 07 & -114.5 & -132.5 & 4.7 & bkgr & -12.3 & 14.0 & 1.5$\times$10$^{4\ast}$ & 120.7 & 110.2 & 20.0 & 2.8$\times$10$^{2\ast}$ & 51.8 & 41.0 & 6.08$\times$10$^{26}$ & 1.07 & 8 & 7.10$\times$10$^{27}$ & (1.89$\pm$0.63)$\times$10$^{28}$ \\
11 & 2013 Apr 11 & -140.7 & -65.4 & -5.4 & 3.4$\times$10$^{4}$ & 65.8 & -11.5 & 1.1$\times$10$^{5}$ & 134.8 & -175.7 & -7.7 & 2.1$\times$10$^{3}$ & 13.6 & 54.1 & 9.41$\times$10$^{26}$ & 1.35 & 8 & 8.47$\times$10$^{28}$ & (1.70$\pm$0.69)$\times$10$^{28}$ \\
12 & 2013 Oct 28 & -141.2 & -50.8 & 4.5 & 9.3$\times$10$^{4}$ & 103.3 & 9.6 & 1.8$\times$10$^{4}$ & 149.5 & -133.6 & 3.7 & 3.4$\times$10$^{3\sharp}$ & -26.9 & 48.1 & 7.87$\times$10$^{26}$ & 9.73 & 8 & 4.41$\times$10$^{29}$ & (8.80$\pm$4.40)$\times$10$^{26}$ \\
13 & 2014 Feb 25 & -159.3 & -38.0 & 15.7 & 1.1$\times$10$^{6}$ & 130.9 & 9.5 & 1.3$\times$10$^{5}$ & 153.7 & -67.9 & 11.8 & 5.9$\times$10$^{4}$ & -47.1 & 29.7 & 3.45$\times$10$^{26}$ & 174.24 & 8 & 3.77$\times$10$^{31}$ & (1.14$\pm$0.52)$\times$10$^{30}$ \\
14 & 2014 Sep 01 & -159.8 & 20.5 & -19.2 & 4.9$\times$10$^{7}$ & 173.4 & -11.7 & 1.8$\times$10$^{5}$ & 167.8 & -16.2 & -15.1 & 4.7$\times$10$^{7}$ & -122.9 & 48.1 & 7.89$\times$10$^{26}$ & 261.00 & 8 & 2.35$\times$10$^{32}$ & (1.99$\pm$0.90)$\times$10$^{30}$ \\
\hline
\end{tabular}
\footnotesize{
($\dagger$) Upper limit.
($\sharp$) Component from a previous SEP event.
($\ast$) Integration interval limited by the onset of a successive SEP event.}
\caption{SEP longitudinal extent at 1 AU. The first two columns report the SEP event number and date. Columns 3--13 give the HEEQ longitude (deg) of STEREO-A/B location (``Loc.''), the longitudinal ($\Delta\beta$, deg) and latitudinal ($\Delta\alpha$, deg) deviation between footpoints of PAMELA or STEREO-A/B at 30 $R_{s}$
and the parent flare location, along with the event-integrated intensities (sr$^{-1}$cm$^{-2}$) above 80 MeV measured by the three spacecraft. Upper limits are provided for the STEREOs;
``bkgr'' indicates a high background from a previous event may have obscured any SEP signal. Columns 14--17 indicate the peak longitude $\beta_{sep}$ (deg) and the standard deviation $\sigma$ (deg) of the SEP spatial distribution derived from the fit of measured intensities (Equation \ref{eq:GaussianProj}), the weighted surface (cm$^{2}$) and the spatial correction factor defined by Equations \ref{eq:Warea} and \ref{eq:SpatialCorr}. 
Columns 18 gives the number of crossings used for the calculation. Finally, column 19--20 report the total number of SEP protons above 500 MeV at 1 AU and the number of protons interacting at the Sun inferred from $\it{Fermi}$/LAT observations by \citet{ref:SHARE2018}}.
\label{tab:all_fit_results}
\end{sidewaystable}

The results of calculation of SEP parameters at 1 AU are shown in Table \ref{tab:all_fit_results}.
The first two columns give the SEP event number and date. 
Columns 3--13 list the longitudinal and latitudinal connection angles between the flare location and the spacecraft magnetic footpoints at 30 $R_{s}$, and the event-integrated intensities measured by the three spacecraft above 80 MeV; for STEREO-A/B upper limits are provided and the spacecraft location is also given; ``bkgr'' indicates a high background from a previous event may have obscured any SEP signal. 
Columns 14--15 display the Gaussian fit parameters (peak longitude and standard deviation), while the values of the weighted spherical surface $S_{J}$ and the spatial correction factor $C_{spa}$ (see Equations \ref{eq:Warea} and \ref{eq:SpatialCorr}) are reported in columns 16 and 17. 
Columns 18 gives the number of crossings used for the calculation. Finally, column 19 lists the numbers of SEP protons above 500 MeV at 1 AU. For comparison, the numbers of protons interacting at the Sun inferred from $\it{Fermi}$/LAT observations by \citet{ref:SHARE2018} are shown in column 20; they include the correction factor accounting for a downward isotropic angular distribution (see Table 3 in \citet{ref:SHARE2018}).

The average $\sigma$ value estimated for the analyzed SEP event sample is $\sim$41 deg, characterized by a large RMS value ($\sim$11 deg). For two SEP events, the one on 2011 August 09 and the GLE event on 2012 May 17, the standard deviation is small ($\sim$21 deg) resulting in a narrow distribution that peaks close to the Earth's magnetic footpoints. On the other hand, the three events on 2011 September 6, 2013 April 11 and 2014 September 1 are very broad, with a $\sigma$ value larger than 50 deg; however, as aforementioned, the integrated intensities measured during the 2011 September 6 event by PAMELA and STEREO-A include a component from the previous SEP event, probably resulting in an overestimate of the spatial extent. 
For the other two events, it can be speculated that the broad distributions are due to significant transport effects, such as cross-field diffusion and IMF co-rotation in combination with the extended SEP source provided by the CME-driven shock.
The rest of the SEP sample is characterized by a standard deviation between 30 and 50 degrees.

Three of the considered events, the 2012 January 23 and 27 events and the 2014 February 25 event, were also investigated by \citet{ref:COHEN2017} at much lower energies ($\leq$10 MeV). 
In general, the standard deviation of the spatial distribution is expected to decrease with increasing energy as a consequence of several acceleration and/or transport related phenomena. This can reasonably explain the relatively small differences regarding the 
2014 February event, for which the derived standard deviations are consistent (33 deg at 10 MeV vs 30 deg above 80 MeV).
On the other hand, they found significantly larger $\sigma$ values for the 2012 January 23 and 27 events (respectively 45 deg and 49 deg at 10 MeV, vs 42 deg and 37 deg above 80 MeV).
As discussed in Section \ref{Deriving Total Proton Numbers from SEP Events}, 
projection effects have a significant impact for high-latitude events such as those in 2012 January, and 
the use of a Gaussian function to fit the particle intensities neglecting the latitude of the SEP propagation axis results in some overestimate of the distribution standard deviation. The correction given by Equation \ref{eq:GaussianProj} likely constitutes a major source of discrepancy between the two calculations. 

\begin{figure}\center
\includegraphics[width=0.8\linewidth]{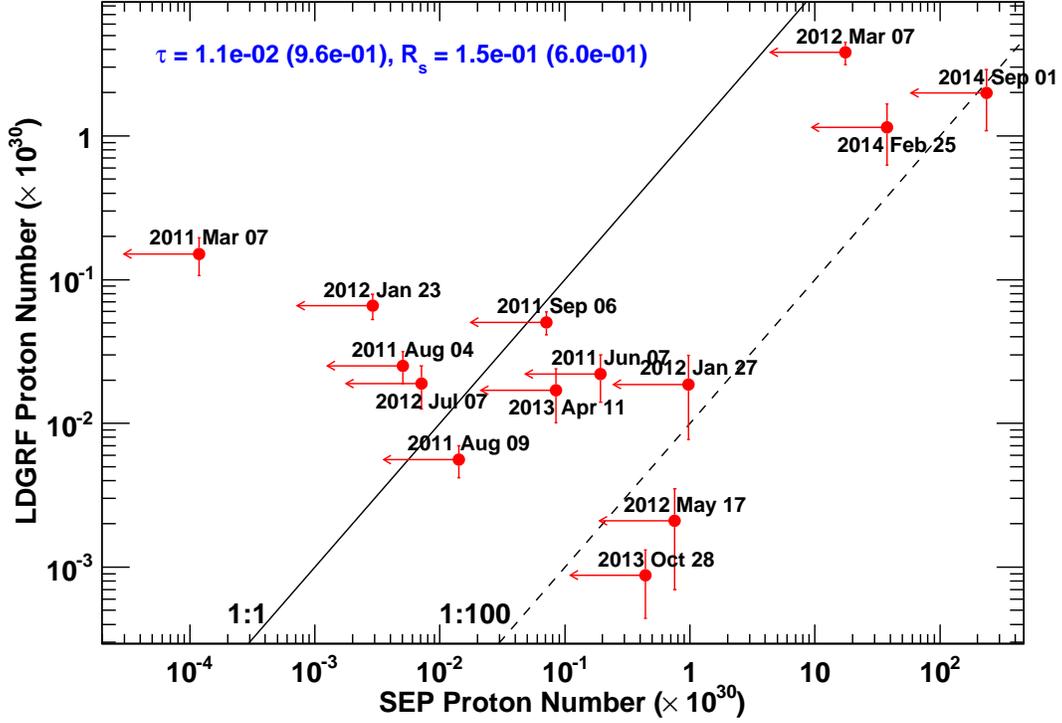}
\caption{Number of protons deduced from $\it{Fermi}$/LAT \citep{ref:SHARE2018} compared with number of protons determined from PAMELA and STEREO-A/B. The solid and the dashed lines mark the one-to-one and the one-to-hundred correspondences, respectively. The Kendall's $\tau$ and the Spearman rank ($R_{s}$) correlation coefficients are also reported, along with corresponding $p$-values.}
\label{fig:Fermi_vs_Pamela_fluences_80MeV}
\end{figure}

\subsection{Comparison between $N_{LDGRF}$ and $N_{SEP}$}
The comparison between proton numbers derived from the $>$100 MeV $\gamma$-ray emission ($N_{LDGRF}$) and SEPs at 1 AU ($N_{SEP}$) is displayed in Figure \ref{fig:Fermi_vs_Pamela_fluences_80MeV}. The solid and the dashed curves mark the one-to-one and the one-to-hundred correspondences, respectively. 
Fermi/LAT proton numbers are from \citet{ref:SHARE2018}, and are corrected for anisotropic effects characterized by the downward proton distribution \citep{ref:MandzhavidzeRamaty1992}.
The vertical error bars include the uncertainties on LDGRF proton numbers from \citet{ref:SHARE2018}. 
For the quantity $N_{SEP}$, upper limits are provided accounting for the assumptions made in the spatial distribution estimate, as discussed earlier. A large scatter among the points can be observed, with no significant correlation as confirmed by the low values of the 
Kendall's $\tau$ and the Spearman rank correlation coefficients, reported on the plot along with corresponding $p$-values (significance level). The $N_{SEP}/N_{LDGRF}$ ratio spans more than five orders of magnitude, ranging from $\sim$7.8$\times$10$^{-4}$ to $\sim$5.0$\times$10$^{2}$, with a mean value of $\sim$78. 

The lowest ratio values are obtained for the 2011 March 7 and the 2012 January 23 SEP events, which were characterized by very soft energy spectra in both PAMELA and STEREO observations. \citet{ref:GOPALSWAMY2018} suggested that the relatively low SEP intensities measured during the 2011 March 7 event are probably due to the poor latitudinal magnetic connectivity. As our calculation accounts for these effects by means of the spatial factor given by Equation \ref{eq:SpatialCorr}, we conclude that the very low $N_{SEP}/N_{LDGRF}$ ratio cannot be explained by connectivity arguments.
Indeed, the $\sim$33 deg latitudinal deviation between PAMELA magnetic foopoints and the flare location for this event is, for instance, comparable to the corresponding angle for the 2012 January 27 event ($\sim$28 deg, see Table \ref{tab:all_fit_results}) which however extended above 500 MeV.
An alternative explanation for the soft spectra is provided by the values of the shock formation height, a key factor in determining the maximum energy of SEPs as the acceleration efficiency strongly depends on the coronal magnetic field strength, which decreases with increasing heliocentric distance \citep{ref:Zank2000}.
Most events shown in Figure \ref{fig:Fermi_vs_Pamela_fluences_80MeV} were accompanied by the type-II radio bursts with high ($>$100 MHz) starting frequency (see, e.g., \citet{ref:MITEVA2017}) suggesting high local plasma densities and thus small ($<$1.6 $R_{s}$) shock formation heights based on the 
empirical formula by \citet{ref:GOPALSWAMY2013}. In contrast, 
using STEREO coronagraphic and EUV observations close to the solar surface,
\citet{ref:GOPALSWAMY2013} estimated a 1.93 $R_{s}$ height for the 2011 March 7 event, whose type-II emission was limited to relatively lower frequencies ($f_{max}$$\sim$50 MHz). A similar height (2 $R_{s}$) was computed by \citet{ref:JOSHI2013} for the 2012 January 23 event; they also proposed that the unusually high $f_{max}$ value ($\sim$200 MHz) associated with the eruption was probably due to the fact that the shock formed in the body of a previous CME.

The highest $N_{SEP}/N_{LDGRF}$ ratio values correspond to the 2012 May 17 GLE event and the 2013 October 28 event, for which the detected $>$100 MeV $\gamma$-ray emission was rather low. 
$N_{SEP}$ is somewhat overestimated for the latter event due to the presence of a previous event, but this alone is insufficient to account for the high $N_{SEP}$/$N_{LDGRF}$ ratio. Similar circumstances in other events also do not produce the large scatter in the values of this ratio.

\begin{figure}\center
\includegraphics[width=0.75\linewidth]{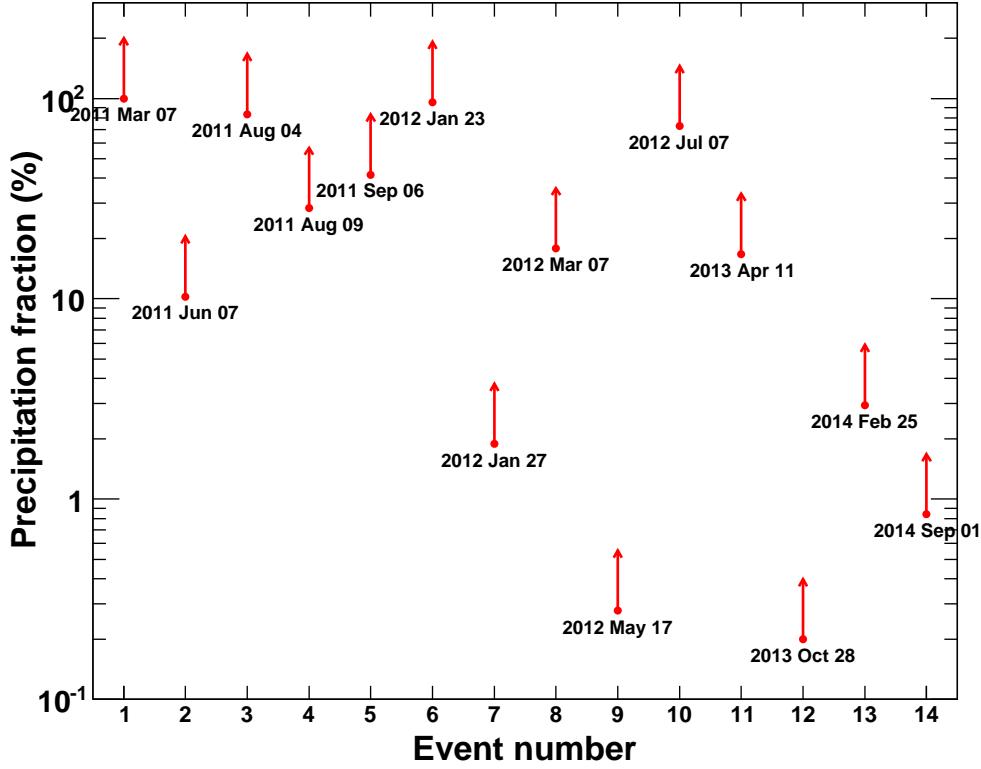}
\caption{Lower limits on the fraction of precipitating protons $N_{LDGRF}/(N_{SEP}+N_{LDGRF})$.}
\label{fig:fraction_time}
\end{figure}

Another way of comparing the proton numbers is to calculate the precipitation fraction, reported in Figure \ref{fig:fraction_time} by event number. This is the fraction $N_{LDGRF}/(N_{LDGRF}+N_{SEP})$ of the total number of protons accelerated (those that escape as SEPs plus those that produce the LDGRFs) that would have to precipitate to account for the LDGRFs. 
For five events only a relatively small fraction of particles ($\lesssim$3\%) are required to precipitate back to the Sun while for the rest of the sample the fraction is larger than 10\%, with three events that require $\gtrsim$83\% to precipitate back to the Sun. We conclude that the large fraction of precipitating protons required to explain the LDGRF emission in several of the events surveyed and the lack of any pattern in the ratio of $N_{LDGRF}$ over the sum of $N_{LDGRF}$ and $N_{SEP}$ pose significant challenges to the CME-driven shock scenario as the source of LDGRF emission.

\subsection{Uncertainties on the SEP propagation axis}
In Section \ref{Deriving Total Proton Numbers from SEP Events}, we assumed that the latitudinal angle of the SEP spatial distribution peak coincides with the latitude of the parent flare ($\alpha_{sep}=\alpha_{flare}$). 
However, the SEP propagation axis can be more closely associated with the related CME direction. The CME may also be launched non-radially above the active region's core or be deflected, for example by coronal holes, with significant implications in terms of latitudinal connectivity \citep{ref:GOPALSWAMYMAKELA2014,ref:GOPALSWAMY2014}.
We estimated the effect of the above approximation on $N_{SEP}$ by comparing the derived SEP distributions with those obtained by using the CME directions from DONKI (see column 11 in Table \ref{tab:sep-events}) in place of flare locations ($\alpha_{sep}=\alpha_{cme}$). The resulting differences in the Gaussian fit parameters ($\beta_{sep}$ and $\sigma$) are both $\lesssim$3.6 deg, while the resulting variations in $N_{SEP}$ are $\lesssim$27\%, with the exception of the 2012 January 27 event, for which the $N_{SEP}$ value is $\sim$47\% larger. This is likely due to the relatively large (13 deg) discrepancy between $\alpha_{cme}$ and $\alpha_{flare}$. The SEP distributions were also compared with those computed by using the CME flux rope directions from \citet{ref:GOPALSWAMY2014,ref:GOPALSWAMY2015}. In this case, the variations in the $\beta_{sep}$ and $\sigma$ parameters are $\lesssim$1.5 deg and $\lesssim$0.6 deg respectively, and a $\sim$15\% maximum difference is found for the corresponding $N_{SEP}$ values. These results suggest that uncertainties in the latitudinal angle of the SEP spatial distribution peak are not expected to affect significantly the comparison between the $N_{SEP}$ and $N_{LDGRF}$ values.

\subsection{Comparison with previous works}
An estimate of the numbers of SEPs at 1 AU has been also provided by \citet{ref:SHARE2018} for some events in common with the present analysis. 
Comparison of these SEP numbers is complicated by the fact that the adopted approaches are significantly different in terms of the derived SEP spatial distribution, transport effects and the datasets studied. In particular, \citet{ref:SHARE2018} assumed, for all the considered SEP events, a Gaussian distribution derived by combining the neutron monitor observations of 57 GLE events. The neutron monitor data for two GLE observations were also used to derive the number of 1 AU crossings, which was found to be much lower ($\overline{N}_{cross}$=2) than those evaluated in this work. 
Furthermore, \citet{ref:SHARE2018} use GOES/HEPAD proton data with a less accurate calibration scheme than that used here.
Moreover, these data were fit with a pure power-law function, neglecting the high-energy spectral rollovers reported by the PAMELA experiment \citep{ref:BRUNO2018}, and this results in a large overestimate of particle intensities above 500 MeV. 
As a consequence, the corresponding SEP proton numbers estimated by \citet{ref:SHARE2018} are, on average, a factor $\sim$20 larger than those evaluated in this work.

\section{Discussion}\label{Discussion}
One of the prominent explanations for the extended duration of LDGRFs is continuous back precipitation of energetic protons accelerated at a CME-driven shock as it propagates outward from the Sun. In this model both the LDGRF and 1 AU SEP signatures are due to protons drawn from the same population. However, because of the strong mirroring effect impeding the transport of protons back to the Sun \citep{ref:Hudson2017}, this scenario would require $N_{SEP}$ to be significantly larger than $N_{LDGRF}$. 
The picture that emerges from our analysis is not consistent with a source rooted in back precipitation from CMEs. For example, 
even with large systematic uncertainties, 
the 2012 May 17 event misses the equality line in Figure \ref{fig:Fermi_vs_Pamela_fluences_80MeV} by a factor of $\sim$10$^{3}$ in $\gamma$-ray producing protons and it differs from the 2012 March 7 event by a factor of $>$10$^{3}$ for similar integrated particle number at 1 AU. 
This could be explained by the likely sporadic and unpredictable magnetic connection between the shock front and the Sun. Nevertheless, it would follow that the intensity-time profile in any given LDGRF would be similarly wildly varying, atypical of most well-measured LDGRFs observed by $\it{Fermi}$/LAT. 

The events to the upper left of the equality are doubly problematic because they represent events where the particle number at the flare exceeds, sometimes considerably, the particle number in space. The implications of this pose a serious challenge to theory.
If the particles above 500 MeV in space were from the same population as those responsible for the $\gamma$-ray emission at the Sun, it would imply that in some cases, more than $\sim$80\% of this population must be extracted
from the acceleration process to produce the $\gamma$ radiation. 
In particular, although the estimate of $N_{SEP}$ is subject to large uncertainties related to multiple injections, the number of protons inferred from the $\it{Fermi}$/LAT detection during the 2011 March 7 event, associated with the second longest duration $\gamma$-ray emission, is more than four orders of magnitude larger. Losses of this magnitude from the shock would certainly quench the particle acceleration process long before particles of sufficient energy and number could be produced by the shock. 
Furthermore, because the shock formation height estimated for the 2011 March 7 and 2012 January 23 events is relatively large ($\sim$2 $R_{s}$, \citet{ref:GOPALSWAMY2013,ref:JOSHI2013}),
the challenge to get particles back from a weakening shock to the Sun is even greater. 

The problem is exacerbated for events such as the 2012 March 7 and the 2014 February 25 events, because the SEP spectra fall off exponentially 
above few hundreds MeV \citep{ref:BRUNO2018} while there are abundant $\gamma$-rays exceeding 1 GeV from the flare \citep{ref:SHARE2018}. Consequently, a particle number plot similar to that for Figure \ref{fig:Fermi_vs_Pamela_fluences_80MeV}, but for higher energies, would push
the points much farther upwards and to the left of an equality line, aggravating the problem. 
Although CME-driven shock protons exhibit a prolonged, delayed and high-energy behavior, we find it difficult to reconcile the back precipitation scenario with the production of the high-energy $\gamma$ radiation at the Sun because it seems to badly fail the total particle number test, implying a separate process, associated with the CME only in a general order-of-magnitude sense, consistent with the trend in the data. 

Complications arise in other ways, in particular from
two $\it{Fermi}$/LAT detections on 2012 October 23 and 2012 November 27, which were not linked to CME eruptions, suggesting that a fast CME is not a necessary requirement of LDGRFs.
In addition, while it has been shown that many bright X-ray flares are not associated with LDGRFs \citep{ref:WINTER2018}, 
there are also counter-examples of very fast front-side CMEs that were not accompanied by
$>$100 MeV $\gamma$-ray emission, such as the two full halo CMEs observed at 10:48 UT on 2011 September 22 (1905 km s$^{-1}$) and at 12:48 UT on 2011 September 24 (2018 km s$^{-1}$)
and the partial halo CME registered at 03:12 UT on 2013 June 21 (1900 km s$^{-1}$).
Similarly, while all reported LDGRFs are associated with impulsive-flare hard X-rays with energies above 100 keV, the latter is not a sufficient condition for the production of LDGRFs \citep{ref:SHARE2018}. Finally, no long-duration $\gamma$-ray observation was reported by $\it{Fermi}$/LAT during the
two front-side eruptions producing $>$500 MeV SEPs
measured by PAMELA on 2012 March 13 and 2012 July 8. 
However, firm conclusions about the absence of $>$100 MeV $\gamma$-ray emission in such events are complicated by the limited duty cycle of LAT when viewing the Sun. 

An alternative to back precipitation is the scenario where particles are accelerated via second-order Fermi mechanism and trapped locally, within extended coronal loops, after which (or concurrently) they diffuse to the denser photosphere to radiate \citep{ref:RYANLEE1991}. 
The favorable conditions can be provided by the magnetic structures
appearing during the gradual phase of two-ribbon flares and CME liftoff, as field-line reconnection gives rise to hot flare loops whose size can exceed 1 $R_{s}$,
creating a system of arches that can persist for several hours.
Based on observations of gyrosynchrotron sources (GS) from the Nancay Radioheliograph (NRH), \citet{ref:GRECHNEV2018} provided evidence that the behind-the-limb flare of 2014 September 1, involved two distinct quasi-static loops, one loop that is associated with an initial HXR flare and another larger loop associated with a second HXR flare. Comparing these observations with simulations, they noted that the time profiles of the GS, HXR, and to some extent the $>$100 MeV $\gamma$-ray emission from $\it{Fermi}$/LAT, of the latter microwave source are consistent with prolonged confinement (and perhaps re-acceleration) of particles injected within a magnetic trap. 

A compelling association of a large, extended loop with $>$100 MeV $\gamma$-ray emission is also provided by recent microwave observations from EOVSA \citep{ref:GARY2018}. During the 2017 September 10 solar eruption, that produced a GLE and also a LDGRF, EOVSA microwave observations identified the footpoints of a large coronal loop, with circular length of $\sim$1.4 $R_{s}$. The microwave emission persisted well into the extended phase of the $>$100 MeV $\gamma$-ray emission. The robust and smooth exponential decay of the 
latter argues for the coronal trap scenario, with spatial and momentum diffusion governing the precipitation of high-energy particles \citep{ref:RYAN2018b}, as seen in every other well measured LDGRF \citep{ref:SHARE2018}. 
This approach 
conveniently de-couples the acceleration of the $\gamma$-producing particles from the acceleration and transport of the SEPs which in turn allows for different spectral shapes as well as different energetic particle productivities, as seen in this analysis. The intrinsic prolonged durations of the $\gamma$-ray signatures from both processes speak to the large requisite spatial scales for both processes and the resulting high energies. Because the loop scenario is local and diffusive in nature, it naturally would produce smooth exponential decays, since no sequence of magnetic connects or disconnects would occur, as would be expected for a propagating large scale feature like a CME.

A problem encountered in wider spread acceptance of this scenario is that although large loop structures ($\gtrsim$1 $R_{s}$) are common, they are often difficult to visualize, not filled with hot plasma or enough 100 keV electrons to be visible in soft X-ray or radio emissions.
For the purpose of accelerating protons, only magnetic turbulence or Alfv\'en waves is necessary with $\delta B/B$ $\sim$ 10\% \citep{ref:RYAN2018a}.
New radio observations can help to place constraints on loop size and the ambient conditions within the loop that will improve future modeling of LDRGFs within the context of the continuous acceleration and trapping scenario.

\section{Summary}\label{Summary}
Taking advantage of the unique high-energy observations from PAMELA, we conducted the first direct comparison of 
the number of interacting protons at the Sun and
the number of SEP protons
in the energy range above the $\sim$300 MeV pion-production threshold. Several key factors contribute to the analysis including the ability to obtain SEP energy spectra up to few GeV, firmly establishing spectral roll-overs for all high-energy SEP events observed \citep{ref:BRUNO2018}. 
In particular we calculate the total number of
$>$500 MeV protons at 1 AU by combining PAMELA and STEREO data with the aid of transport simulations, and compare it with the number of high-energy protons at the Sun, as deduced from $\it{Fermi}$/LAT data \citep{ref:SHARE2018}. The results of our analysis shows that the two proton numbers are uncorrelated 
such that their ratio spans more than five orders of magnitude. The lack of 
correlation, and in particular several extreme cases where the number of protons required to account for LDGRFs far exceeds the number of SEP protons, suggests that the LDGRF emission is probably not due to the back precipitation of particles accelerated at CME-driven shocks. 
Moreover, as demonstrated by the two LDGRF events occurring on 2012 October 23 and November 27, the association with fast CMEs does not appear to be a necessary requirement for high-energy $\gamma$-ray emission. In fact, while bright flares and impulsive $>$100 keV hard X-rays are not sufficient conditions for LDGRFs, there also several counter-examples of fast halo CMEs that were not accompanied by $>$100 MeV $\gamma$-ray emission, though the limited 
exposure of the LAT instrument complicates the interpretation. 
An alternative explanation for LDGRFs based on continuous particle acceleration and trapping within large coronal structures that are not causally connected to the CME shock is discussed, and new remote observations of these loops, such as those provided by EOVSA, may help to constrain the role of such acceleration in producing LDGRF emission.

\acknowledgments
The authors thank the ACE, GOES and STEREO teams for making their data publicly available.
G.~A.~de N. acknowledges support from Fermi/GI grant NNH10ZDA001N-FERMI, NASA/HSR grant NNH13ZDA001N-HSR, and NASA/ISFM grant HISFM18. 
A.~B. acknowledges support by an appointment to the NASA postdoctoral program at the NASA Goddard Space Flight Center administered by Universities Space Research Association under contract with NASA. 
S. D. acknowledges support from the UK Science and Technology Facilities Council (STFC) (grant ST/R000425/1).
J.~G. acknowledges support from NASA grant NNX15AJ 71G and NSF grant 1735422.
I.~G.~R. acknowledges support from NASA Living With a Star grant NNG06EO90A.

\end{document}